\documentclass[12pt]{article}
\usepackage{a4wide,amssymb,cite}
\pdfoutput=1

\usepackage{a4wide,amssymb,graphicx}
\usepackage{epsfig}
\usepackage[usenames,dvipsnames]{color}
\usepackage{slashed}
\newcommand{\be}{\begin{equation}}
\newcommand{\ee}{\end{equation}}
\newcommand{\bea}{\begin{eqnarray}}
\newcommand{\eea}{\end{eqnarray}}

\def\circa#1{\,\raise.3ex\hbox{$#1$\kern-.75em\lower1ex\hbox{$\sim$}}\,}

\begin{document}

\begin{titlepage}
%
%


%

\begin{centering}
\vspace{1cm}
{\Large {\bf Unitary inflaton as decaying dark matter} } \\

\vspace{1.5cm}

{\bf Soo-Min Choi$^{1,\dagger}$, Yoo-Jin Kang$^{1,\ddagger}$, Hyun Min Lee$^{1,2,*}$ and Kimiko Yamashita$^{3,\#}$ }
\vspace{.5cm}

{\it $^1$Department of Physics, Chung-Ang University, Seoul 06974, Korea.} 
\vspace{0.1cm} \\
{\it $^2$School of Physics, Korea Institute for Advanced Study, Seoul 02455, Korea.}
\vspace{0.1cm}  \\
{\it $^3$Department of Physics, National Tsing Hua University, Hsinchu, Taiwan 300.}
\\

\end{centering}
\vspace{2cm}

\begin{abstract}
\noindent
We consider the inflation model of a  singlet scalar field (sigma field) with both quadratic and linear non-minimal couplings where unitarity is ensured up to the Planck scale. We assume that a $Z_2$ symmetry for the sigma field is respected by the scalar potential in Jordan frame but it is broken explicitly by the linear non-minimal coupling due to quantum gravity. 
We discuss the impacts of the linear non-minimal coupling on various dynamics from inflation to low energy, such as a sizable tensor-to-scalar ratio, a novel reheating process with quartic potential dominance, and suppressed physical parameters in the low energy, etc.
In particular, the linear non-minimal coupling leads to the linear couplings of the sigma field to the Standard Model through the trace of the energy-momentum tensor in Einstein frame.  
Thus, regarding the sigma field as a decaying dark matter, we consider the non-thermal production mechanisms for dark matter from the decays of Higgs and inflaton condensate and show the parameter space that is compatible with the correct relic density and cosmological constraints.

\end{abstract}

\vspace{2cm}

 \begin{flushleft}
 $^\dagger$Email: soominchoi90@gmail.com \\
 $^\ddagger$Email: yoojinkang91@gmail.com \\
$^*$Email: hminlee@cau.ac.kr \\
$^\#$Email: kimiko.y.phys@gmail.com 
 \end{flushleft}

\end{titlepage}

\section{Introduction}

Measurements of anisotropies of Cosmic Microwave Background (CMB) provide important clues to the early Universe after Big Bang, such as inflation, dark matter and dark energy. In particular, it has been shown that the observed CMB spectrum \cite{planck,planck2018} is consistent with the predictions from the slow-roll inflation with a single canonical scalar field, the so called inflaton.  
However, what causes the inflation is unknown, although some of the early proposed inflation models including quartic or quadratic inflaton potentials have been now disfavored.

Higgs inflation \cite{Higgsinf} has been proposed as an economic implementation of inflation in particle physics, just with a single non-minimal coupling of Higgs field to gravity, so it has drawn a lot of attention from both particle physics and cosmology communities. Some time after the proposal, it was also noticed that a large non-minimal coupling necessary for a successful Higgs inflation causes unitarity problem much below the Planck scale \cite{unitarity}. However, Higgs inflation can be saved under the assumption that new physics entering at unitarity scale respects the approximate scale symmetry \cite{fedor} or due to extra degrees of freedom fully recovering the unitarity up to the Planck scale \cite{gian,espinosa,gian2,R2inflation}.

Recently, a new proposal for unitarizing Higgs inflation with a light inflaton, dubbed the sigma field $\sigma$, has been made by one of the authors \cite{hmlee}. In this case, the inflaton carries both quadratic and linear non-minimal couplings. As a result, we can keep the flat direction for inflation due to a large quadratic non-minimal coupling and, at the same time, unitarity scale is restored up to the Planck scale due to the field-independent rescaling of the inflaton field due to the linear non-minimal coupling. In this framework, the sigma field mass can take any value below the unitarity scale of the original Higgs inflation such that we can recover the Higgs inflation in the effective theory, but with a sizable correction to the tensor-to-scalar ratio at tree level. 

In this article, we investigate  the impacts of the linear non-minimal coupling on various dynamics from inflation to low energy phenomena such as reheating and inflaton couplings, in connection to unitarity scale and inflationary predictions. 
The $Z_2$ symmetry for $\sigma$ is respected by the inflaton potential in Jordan frame but it is broken explicitly by the linear non-minimal coupling. Then, the inflaton has novel couplings to the SM through the trace of the energy-momentum tensor with a suppression of the Planck scale in Einstein frame. In this case, we pursue the possibility that the inflaton can be a decaying dark matter (DM). To this, we consider the non-thermal production mechanisms for dark matter from the decays of the SM Higgs boson and the inflaton condensate. As a result, we show the parameter space for a decaying dark matter that is consistent with the correct relic density and cosmological constraints.
There has been a recent proposal of axion-like inflation where the inflaton makes a decaying dark matter too \cite{ALP}.

The paper is organized as follows.
We begin with a model description for the sigma field inflation and discuss the inflationary dynamics along the flat direction in the system with sigma and Higgs fields. Then, we continue to consider the reheating dynamics with a novel quartic potential and identify the possible reheating temperature depending on the mixing quartic coupling. Next we show how the unitarity is restored up to the Planck scale due to a sizable linear non-minimal coupling and identify the low energy parameters in the potential and inflation couplings to the SM through the trace of the energy-momentum tensor. 
As a result, the decay branching ratios of the inflaton are shown for heavy or light inflaton cases. Then, we show the parameter space for inflaton dark matter that is consistent with the correct relic density and the Big Bang Nucleosynthesis (BBN), CMB and large-scale structure bounds. 
Finally, conclusions are drawn.
There is one appendix dealing with the details on inflaton decay rates.

\section{Model}

We consider the inflation model with a real scalar field $\sigma$ as a simple extension of the Standard Model (SM), with the corresponding Lagrangian \cite{hmlee} being
\bea
\frac{\cal L}{\sqrt{-g}} &=& -\frac{1}{2} \Omega(\sigma,H) R + \frac{1}{2}(\partial_\mu\sigma)^2+|D_\mu H|^2 - V(\sigma,H) \nonumber \\
&&-\frac{1}{4g^2} V_{\mu\nu} V^{\mu\nu} +{\bar\psi} i\gamma^\mu \Big(D_\mu+\frac{1}{2}\omega^{ab}_\mu \sigma_{ab}\Big) \psi - (y H{\bar \psi}_L \psi_R +{\rm h.c.})
\label{sigma-jordan}
\eea
where the frame function and the scalar potential are given by
\bea
\Omega(\sigma,H) &=& 1+\xi_1 \sigma +\xi_2\sigma^2 +2\xi_H |H|^2,   \label{frame} \\
V(\sigma,H) &=&V_0+ \frac{1}{2} m^2_\sigma \sigma^2  +\frac{1}{4} \lambda_\sigma \sigma^4 +\frac{1}{2} \lambda_{\sigma H}\sigma^2  |H|^2 +m^2_H |H|^2  +\lambda_H |H|^4. \label{jpot}
\eea
We note that $H$ is the SM Higgs doublet, $V_{\mu\nu}$ is to collectively describe the field strength tensors for the SM gauge bosons, and $\psi$ are the SM fermions, and $D_\mu H$ and $D_\mu\psi$ are covariant derivatives, $\sigma_{ab}=\frac{1}{4}[\gamma_a,\gamma_b]$, and $V_0$ is a constant term which is chosen to set the cosmological constant in the vacuum to zero. 
Here, we assume that the sigma field is odd under a $Z_2$ symmetry, i.e. $\sigma\rightarrow -\sigma$, that is respected by the scalar potential but broken only due to the linear non-minimal coupling $\xi_1$ in quantum gravity. As will be discussed later,  it is still possible to  make the inflaton as a decaying dark matter if light enough, even in the presence of the violation of the $Z_2$ symmetry.

We note that the $Z_2$ symmetry gets restored in the limit of a vanishing $\xi_1$, so it is natural to introduce the approximate $Z_2$ symmetry in the low energy.  As will be shown in Sections 5 and 6,  the $Z_2$ breaking is communicated to the SM via gravitational interactions, thus it appears as higher dimensional interactions with suppression scales larger than or equal to the Planck scale in Einstein frame. Then, we can regarding our setup as an effective theory below the Planck scale. Therefore, as far as higher dimensional operators  are suppressed by the Planck scale at least, our later discussion based on the $Z_2$ breaking non-minimal coupling holds.

When the sigma field is heavier than electroweak scale, it is too short-lived to be a dark matter candidate.
In this case, after integrating out a heavy sigma field with $\lambda_{\sigma H}<0$, we obtain a Higgs effective theory with the effective frame function and scalar potential \cite{hmlee}, given by
 \bea
 \Omega_{\rm eff}&=&1-\frac{\xi_2 m^2_\sigma}{\lambda_\sigma}+\xi_1\sqrt{\frac{-m^2_\sigma-\lambda_{\sigma H}|H|^2}{\lambda_\sigma}}+2\xi_{H,{\rm eff}} |H|^2, \\ 
  V_{\rm eff}&=& V_{0,{\rm eff}}+ m^2_{H,{\rm eff}} |H|^2 + \lambda_{H,{\rm eff}} |H|^4
 \eea
 with
 \bea
 V_{0,{\rm eff}}&=&V_0-\frac{m^4_\sigma}{4\lambda_\sigma}, \\
 m^2_{H,{\rm eff}}&=&m^2_H-\frac{\lambda_{\sigma H}}{2\lambda_\sigma}\, m^2_\sigma, \\
 \xi_{H,{\rm eff}} &\equiv&\xi_H-\frac{\lambda_{\sigma H}\xi_2}{2\lambda_\sigma},  \\
 \lambda_{H,{\rm eff}}&\equiv &\lambda_H-\frac{\lambda^2_{\sigma H}}{4\lambda_\sigma}. 
 \eea
 Therefore, the effective Higgs quartic coupling $ \lambda_{H,{\rm eff}}$ gets a tree-level shift due to the scalar threshold, curing the vacuum instability problem in the SM\cite{oleg-vsb,sthreshold}.
 Moreover, a large positive effective non-minimal coupling $ \xi_{H,{\rm eff}} $ for the Higgs field can be obtained and the effective frame function also contains a non-analytic form of the non-minimal coupling to gravity for the Higgs field, being proportional to the linear non-minimal coupling for the sigma field.
 However, we will fully take into account the sigma field in our later discussion and focus on the case with a light sigma field.

For $\xi_2, \xi_H>0$, in order to maintain the effective Planck mass squared in Jordan frame to be positive during the cosmological evolution,
we impose the condition for stable gravity \cite{hmlee} as
 \be
 \xi^2_1<4 (\xi_2+\xi_H \tau^2) \label{stablegravity}
 \ee
with $\tau^2=\frac{2|H|^2}{\sigma^2}$. Then, eq.~(\ref{stablegravity}) leads to the upper bound on the linear non-minimal coupling $\xi_1$ for stable gravity in the entire field space.  We will take this into account in the later discussion on inflationary dynamics.

Choosing the Higgs doublet in unitary gauge as $H^T=(0,\phi)/\sqrt{2}$ and performing the metric rescaling by $g_{\mu\nu}=g^E_{\mu\nu}/\Omega$  with $\Omega=1+\xi_1 \sigma+\xi_2\sigma^2+\xi_H \phi^2$,  
we get the Einstein frame Lagrangian of our model as 
\bea
\frac{{\cal L}_E}{\sqrt{-g_E}}&=& -\frac{1}{2} R(g_E) +\frac{1}{2\Omega}(\partial_\mu{\sigma})^2 +\frac{3}{4}(\partial_\mu\ln \Omega)^2+\frac{1}{2\Omega}\Big((\partial_\mu \phi)^2+ \delta_V\,m^2_{V,0}\,  \frac{\phi^2}{v^2} V_\mu V^\mu\Big) \nonumber \\
&& - V_E(\sigma,\phi)-\frac{1}{4g^2} V_{\mu\nu} V^{\mu\nu} +{\bar f} i\gamma^\mu \Big(D_\mu+\frac{1}{2}\omega^{ab}_\mu \sigma_{ab}\Big) f - \Omega^{-1/2}\,\frac{m_{f,0}}{v}\, \phi{\bar f}  f  
\label{sigma-ein} 
\eea
where $m_{f,0}, m_{V,0}$ are SM fermion and electroweak gauge boson masses, independent of the sigma field, and $\delta_V=1(2)$ for $V=Z(W)$ bosons, and the Einstein frame potential is given by
\bea
V_E({\sigma},\phi)= \frac{1}{\Omega^2}\, V(\sigma,\phi) \label{einstein-pot}
\eea
with
\bea
V({\sigma},\phi)&=&V_0+\frac{1}{2} m^2_\sigma \sigma^2 +\frac{1}{4} \lambda_\sigma \sigma^4 +\frac{1}{4} \lambda_{\sigma H}\sigma^2  \phi^2 +\frac{1}{2} m^2_H \phi^2  +\frac{1}{4}\lambda_H \phi^4.
\eea
Here, we note that the SM fermions are rescaled by $f=\Omega^{-3/4} \psi$ for canonical kinetic terms in eq.~(\ref{sigma-ein}) and the form of covariant kinetic terms for fermions is unchanged under the Weyl transformations of the metric and fermions. 

From the following,
\be
\partial_\mu\ln\Omega=\frac{1}{\Omega}\,\Big[(\xi_1+2\xi_2\sigma)\partial_\mu\sigma+2\xi_H \phi\, \partial_\mu \phi \Big],
\ee
and the Einstein frame Lagrangian given in eq.~(\ref{sigma-ein}),
the scalar kinetic terms in Einstein frame can be rewritten as
\bea
\frac{{\cal L}_{\rm kin}}{\sqrt{-g_E}} &=& \frac{1}{2\Omega^2}\, \Big(1+\frac{3}{2}\xi^2_1+(1+6\xi_2)(\xi_1\sigma+\xi_2\sigma^2)+\xi_H \phi^2 \Big)(\partial_\mu\sigma)^2 \nonumber \\
&&+\frac{1}{2\Omega^2}\, \Big(1+\xi_1\sigma+\xi_2\sigma^2+\xi_H (1+6\xi_H) \phi^2 \Big) (\partial_\mu \phi)^2 \nonumber \\
&& +\frac{3}{\Omega^2}\, \xi_H (\xi_1+2\xi_2\sigma) \phi\, \partial_\mu\sigma \partial^\mu \phi. \label{kinterm}
\eea
We will make use of the above form of the kinetic terms for our later discussion on inflationary dynamics and unitarity scales in the true vacuum. 

Furthermore, from eq.~(\ref{sigma-ein}), we also note that the inflaton couplings to the SM in Einstein frame can be read from
\bea
\frac{{\cal L}_{{\rm int}}}{\sqrt{-g_E}}=  \frac{1}{2\Omega}\, (\partial_\mu \phi)^2-\frac{1}{\Omega^2}\, V-\frac{m_{f,0}}{\Omega^{1/2}}\,\frac{\phi}{v}\,{\bar f}f +\frac{1}{2\Omega}\, \delta_V\, m^2_{V,0}\,\frac{\phi^2}{v^2}\, V_\mu V^\mu.  \label{sint}
\eea  
The above interaction Lagrangian will be useful for discussing the reheating dynamics and inflation couplings at low energy in the later sections.

\section{Inflationary dynamics}

We consider the inflationary dynamics in our model with sigma and Higgs fields along the flat direction and discuss the details of the vacuum structure during inflation. After obtaining the effective potential for a single inflaton, we show the differences from the usual inflation with quadratic non-minimal couplings only.

\subsection{Inflaton dynamics along the flat direction}

Taking $\xi_1 \sigma+\xi_2\sigma^2+\xi_H \phi^2 \gg 1$ during inflation, 
we get $\Omega\approx\xi_1 \sigma+\xi_2\sigma^2+\xi_H \phi^2 $ and introduce a  new set of fields \cite{hmlee} by 
\bea
e^{\frac{2}{\sqrt{6}} \chi}&=&\xi_1 \sigma+\xi_2\sigma^2+\xi_H \phi^2, \label{chi} \\
\tau&=& \frac{\phi}{\sigma}. \label{tau}
\eea
Then, from the approximate relation between $\sigma$ and redefined fields, $\chi$ and $\tau$,  given by 
\bea
\sigma\approx \frac{e^{\frac{1}{\sqrt{6}}\chi}}{(\xi_2+\xi_H \tau^2)^{1/2}} \left(1-\frac{\hat R}{2} \,e^{-\frac{1}{\sqrt{6}}\chi}+\frac{{\hat R}^2}{8}\, e^{-\frac{2}{\sqrt{6}}\chi}\right)
\eea
with 
\be
{\hat R}\equiv \frac{\xi_1}{(\xi_2+\xi_H \tau^2)^{1/2}},
\ee
the scalar potential in Einstein frame\footnote{We have corrected the typo in the last term of the scalar potential in Ref.~\cite{hmlee}: $(2+{\hat R}^2)\rightarrow 2(1-{\hat R}^2)$. } becomes
\bea
V_E(\chi,\tau)&\approx & \frac{1}{4} (\lambda_H \tau^4+\lambda_{\sigma H}\tau^2+\lambda_\sigma)\Big(1+e^{\frac{2}{\sqrt{6}} \chi}\Big)^{-2} \sigma^4 \nonumber \\
&\approx&V_I(\tau) \left(1-2 {\hat R} \,e^{-\frac{1}{\sqrt{6}}\chi} -2(1-{\hat R}^2) \, e^{-\frac{2}{\sqrt{6}}\chi} \right)  \label{inflapot}
\eea
with 
\bea
V_I(\tau)\equiv \frac{\lambda_H \tau^4+\lambda_{\sigma H}\tau^2+\lambda_\sigma}{4(\xi_2+\xi_H \tau^2)^2}. \label{taupot}
\eea
Thus, the ratio of the fields is determined dominantly by the minimization of $V_I$ with respect to $\tau$. 
We note that for $\tau=0$, i.e. for zero $\phi$ during inflation, ${\hat R}$ is identical to $R$, that will appear in the unitarity scales in Section 5. We note that the value of ${\hat R}$ is constrained to $0\leq {\hat R}<2$ for stable gravity \cite{hmlee}, as discussed for eq.~(\ref{stablegravity}).

Now we discuss the scalar kinetic terms in Einstein frame and check the consistency of the inflaton identification in the above discussion. 
First, we can rewrite eq.~(\ref{kinterm}) in terms of the shifted sigma field, ${\bar\sigma}=\sigma+\xi_1/(2\xi_2)$, as
\bea
\frac{{\cal L}_{\rm kin}}{\sqrt{-g_E}} &=& \frac{1}{2\Omega^2}\, \Big(1-\frac{\xi^2_1}{4\xi_2}+\xi_2 (1+6\xi_2){\bar\sigma}^2+\xi_H \phi^2 \Big)(\partial_\mu{\bar\sigma})^2 \nonumber \\
&&+\frac{1}{2\Omega^2}\, \Big(1-\frac{\xi^2_1}{4\xi_2}+\xi_2{\bar\sigma}^2+\xi_H (1+6\xi_H) \phi^2 \Big) (\partial_\mu \phi)^2 \nonumber \\
&& +\frac{6}{\Omega^2}\, \xi_H \xi_2\,{\bar\sigma}\, \phi\, \partial_\mu{\bar\sigma} \partial^\mu \phi. \label{kin00}
\eea
Here, during inflation, we can ignore $1-\frac{\xi^2_1}{4\xi_2}$ in eq.~(\ref{kin00}) so the kinetic terms are essentially the same as in the sigma field inflation without the linear non-minimal coupling, although there is a significant difference in the vacuum as will be shown in the later sections.

In the basis of $\chi$ and $\tau$, the Einstein-frame kinetic terms in eq.~(\ref{kin00}) are generically non-diagonal \cite{higgsportal,2HDM,hmlee}. Thus, we have to choose another basis with $\rho$, instead of $\chi$, as follows,
\bea
\rho^2= (1+6\xi_2){\bar\sigma}^2 + (1+6\xi_H)\phi^2.
\eea
This is the Noether current of scale symmetry \cite{ross,scale}, which is approximately respected during inflation because $\rho$ is close to a constant value up to small slow-roll parameters.
Redefining the scalar fields in terms of $\rho$ and $\theta$ \cite{dilaton,scale} as 
\bea
{\bar\sigma}&=& \frac{1}{\sqrt{1+6\xi_2}}\, \rho \cos\theta, \\
\phi&=&\frac{1}{\sqrt{1+6\xi_H}}\, \rho \sin\theta,
\eea
we find that the above Einstein-frame kinetic terms in eq.~(\ref{kin00}) become diagonal,
\bea
\frac{{\cal L}_{\rm kin}}{\sqrt{-g_E}}
&\approx& \left(\frac{\xi_2\cos^2\theta}{1+6\xi_2}+\frac{\xi_H\sin^2\theta}{1+6\xi_H} \right)^{-1} \frac{(\partial_\mu\rho)^2}{2\rho^2} \nonumber \\
&&+\frac{1}{2} \Big(\xi_2\sqrt{\frac{1+6\xi_H}{1+6\xi_2}}\cos^2\theta+\xi_H \sqrt{\frac{1+6\xi_2}{1+6\xi_H}} \sin^2\theta\Big)^{-2}(\xi_2 \cos^2\theta+\xi_H \sin^2\theta) (\partial_\mu\theta)^2 \nonumber \\
&=& \frac{1+6\xi_2+(1+6\xi_H)\tau^2}{\xi_2+\xi_H\tau^2}\,   \frac{(\partial_\mu\rho)^2}{2\rho^2} \nonumber \\
&&+\frac{\xi_2(1+6\xi_2)\xi_H(1+6\xi_H)\tau^2}{1+6\xi_2+(1+6\xi_H)\tau^2}\frac{(\partial_\mu \tau)^2}{2(\xi_2+\xi_H\tau^2)^2}. \label{kin2}
\eea 
This is a more convenient form for discussing the inflaton effective potential with the $\tau$ field decoupled in the next section.

\subsection{Effective action for inflaton}

After stabilization of $\tau$ from the scalar potential $V_I(\tau)$ in eq.~(\ref{taupot}), we get four different vacua for  $\tau$ during inflation and the corresponding minimum conditions \cite{higgsportal}, in the following,
\begin{eqnarray}
&& (1):~\tau = \sqrt{ 
  \lambda_\sigma \xi_H - \lambda_{\sigma H} \xi_2/2  \over 
   \lambda_H \xi_2 - \lambda_{\sigma H} \xi_H /2   }   \;: ~ \lambda_H \xi_2 - \lambda_{\sigma H} \xi_H/2 >0~,~
    \lambda_\sigma \xi_H - \lambda_{\sigma H} \xi_2/2 >0~, \nonumber\\
&&(2): ~\tau=0  \;:~\lambda_H \xi_2 - \lambda_{\sigma H} \xi_H/2 >0~,~
   \lambda_\sigma \xi_H - \lambda_{\sigma H} \xi_2/2 <0~, \nonumber\\
&&(3): ~\tau=\infty  \;:~  \lambda_H \xi_2 - \lambda_{\sigma H} \xi_H/2 <0~,~
   \lambda_\sigma \xi_H - \lambda_{\sigma H} \xi_2/2 >0~,\nonumber\\
&&(4): ~\tau=0,\infty  \;:~ \lambda_H \xi_2 - \lambda_{\sigma H} \xi_H/2 <0~,~
    \lambda_\sigma \xi_H - \lambda_{\sigma H} \xi_2/2 <0~.\label{taumin}
\end{eqnarray}
Then, there is a unique vacuum for $\tau$  in the first three cases, and the vacuum energy in each case given \cite{higgsportal} by
\begin{eqnarray}
(1):&& ~V_I=\frac{1}{4}\frac{\lambda_H\lambda_\sigma-\lambda^2_{\sigma H}/4}{\lambda_\sigma \xi^2_H+\lambda_H\xi^2_2-\lambda_{\sigma H} \xi_H \xi_2},  \nonumber \\
(2):&& ~V_I=\frac{\lambda_\sigma}{4\xi^2_2},  \nonumber \\
(3):&& ~V_I= \frac{\lambda_H}{4\xi^2_H},  \nonumber \\
(4):&& ~V_I=\frac{\lambda_\sigma}{4\xi^2_2}\quad{\rm or}\quad \frac{\lambda_H}{4\xi^2_H},    
\label{taumin-a}
\end{eqnarray}
whereas there are two local minima  in the last case (4), with the same vacuum energy as given in the cases (2) and (3), respectively.

In the case with $\xi_2\gg \xi_H={\cal O}(1)$ and quartic couplings of order unity,  the conditions for the inflation vacua (\ref{taumin})  become
\begin{eqnarray}
&&(1): ~\tau = \sqrt{ - \frac{\lambda_{\sigma H}}{2\lambda_H}  }   \;: ~\lambda_H >0~, ~\lambda_{\sigma H} <0~, \nonumber\\
&&(2): ~\tau=0  \;:~\lambda_H >0~,~
    \lambda_{\sigma H}  >0~, \nonumber\\
&&(3): ~\tau=\infty  \;:~  \lambda_H <0~,~
\lambda_{\sigma H}  <0~,\nonumber\\
&& (4):~\tau=0,\infty  \;:~ \lambda_H <0~,~
   \lambda_{\sigma H}  >0~.\label{taumin2}
\end{eqnarray}
In the first two cases, we need the Higgs quartic coupling to be positive during inflation: the former is the sigma-Higgs mixed inflation and the latter is the pure sigma inflation.  
In the third case, as the Higgs quartic coupling is required to be negative as $\lambda_H<0$, $V_I<0$, so it is not possible to get a dS vacuum for inflation. Finally, in the fourth case, even for $\lambda_H<0$, the inflation could be driven by the sigma field at the metastable vacuum with $\tau=0$ so it could lead to a viable cosmology with correct electroweak symmetry breaking at low energy. But, $\tau=\infty$ is not a valid option because the vacuum energy during inflation is negative, i.e. $V_I=\frac{\lambda_H}{4\xi_H^2}<0$.
The vacuum energy  (\ref{taumin-a}) for the viable inflation is given by
\bea
(1):&& ~V_I=\frac{1}{4\xi^2_2} \Big(\lambda_\sigma-\frac{\lambda^2_{\sigma H}}{4\lambda_H} \Big), \nonumber \\
(2), \, (4):&& ~V_I=\frac{\lambda_\sigma}{4\xi^2_2}.
\label{taumin-aa}
\eea

Therefore, for $\xi_2\gg 1$ and $\theta$ (or $\tau$) stabilized at $\theta_0$ (or $\tau_0$), the Einstein-frame kinetic terms in eq.~(\ref{kin2}) become simplified to
\bea
\frac{{\cal L}_{\rm kin}}{\sqrt{-g_E}}&\approx& \frac{1+6\xi_2+(1+6\xi_H)\tau^2_0}{\xi_2+\xi_H\tau^2_0}\, \frac{(\partial_\mu\rho)^2}{2\rho^2} +\frac{(\partial_\mu \tau)^2}{2\xi_2} \nonumber \\
&=& \frac{1}{2}(\partial_\mu \chi)^2 +\frac{(\partial_\mu \tau)^2}{2\xi_2}.  \label{kintermapprox}
\eea
Here, we note that since
\bea
\Omega&\approx&(\xi_2 {\bar \sigma}^2 + \xi_H  \phi^2)\Big|_{\theta=\theta_0} = \left(\frac{\xi_2\cos^2\theta_0}{1+6\xi_2}+\frac{\xi_H\sin^2\theta_0}{1+6\xi_H} \right)\rho^2= e^{\frac{2}{\sqrt{6}}c\chi},
\eea
where
\bea
c^2=\frac{6(\xi_2+\xi_H\tau^2_0)}{1+6\xi_2+(1+6\xi_H)\tau^2_0}\approx 1, \quad \xi_2\gg 1.
\eea
the above result with the canonical inflaton field $\chi$ is consistent with eq.~(\ref{chi}).

In summary, from eqs.~(\ref{kintermapprox}) and (\ref{inflapot}), the approximate Einstein-frame Lagrangian for inflation is given \cite{hmlee} by
\bea
\frac{{\cal L}_E}{\sqrt{-g_E}}=-\frac{1}{2}R(g_E) +\frac{1}{2} (\partial_\mu \chi)^2 +\frac{(\partial_\mu \tau)^2}{2\xi_2}- V_E(\chi,\tau) \label{infeft}
\eea
with
\be
V_E(\chi,\tau)= V_I(\tau) \left(1-2 {\hat R} \,e^{-\frac{1}{\sqrt{6}}\chi} -2(1-{\hat R}^2) \, e^{-\frac{2}{\sqrt{6}}\chi} \right). 
\ee
We note that the physical mass for the $\tau$ field is rescaled by the non-minimal coupling $\xi_2$ to $\sim  \sqrt{ \xi_2 V_I}\sim \sqrt{\xi_2}\, H_I$, which is much larger than the Hubble scale $H_I$ during inflation, so we can safely ignore the dynamics of the $\tau$ field for the inflationary dynamics.

\subsection{Inflationary predictions}

From the effective inflaton Lagrangian in eq.~(\ref{infeft}),  the slow-roll parameters during inflation are given approximately by
\bea
\epsilon&=& \frac{1}{3} e^{-\frac{2}{\sqrt{6}}\chi} \Big( {\hat R}^2 + 4 {\hat R}\, e^{-\frac{1}{\sqrt{6}}\chi} +4(1+3{\hat R}^2-{\hat R}^4) \, e^{-\frac{2}{\sqrt{6}}\chi}   \Big), \label{epsilon}\\
\eta&=&- \frac{1}{3} e^{-\frac{1}{\sqrt{6}}\chi}\Big({\hat R}+2(2-{\hat R}^2) \, e^{-\frac{1}{\sqrt{6}}\chi}+{\hat R}(10-6{\hat R}^2) \, e^{-\frac{2}{\sqrt{6}}\chi}\Big). \label{eta}
\eea
As a result, the spectral index is given by
\bea
n_s &=&1-6\epsilon_*+ 2\eta_* \nonumber \\
&=& 1-\frac{2}{3}\, e^{-\frac{1}{\sqrt{6}}\chi_*} \Big(  {\hat R}+(4+{\hat R}^2) e^{-\frac{1}{\sqrt{6}}\chi_*}  \Big)
\eea
where $*$ denotes the evaluation of the slow-roll parameters, (\ref{epsilon}) and (\ref{eta}), at horizon exit.
The tensor-to-scalar ratio is also given by $r=16\epsilon_*$ with eq.~(\ref{epsilon}) at horizon exit.
We note that the measured spectral index and the bound on the
tensor-to-scalar ratio are given by $n_s=0.9670\pm 0.0037$ and $r < 0.07$ at $95\%$ C.L., respectively, from Planck 2018 (TT, TE, EE + low E + lensing + BK14 + BAO) \cite{planck2018}, as compared to $n_s=0.9652\pm 0.0047$ and $r < 0.10$ at $95\%$ C.L. in Planck 2015 (TT, TE, EE + low P) \cite{planck}. Thus, the experimental errors in the spectral index from Planck 2018 combination are reduced a bit but the central value of the spectral index is consistent with the one from Planck 2015.

\begin{figure}
  \begin{center}
    \includegraphics[height=0.42\textwidth]{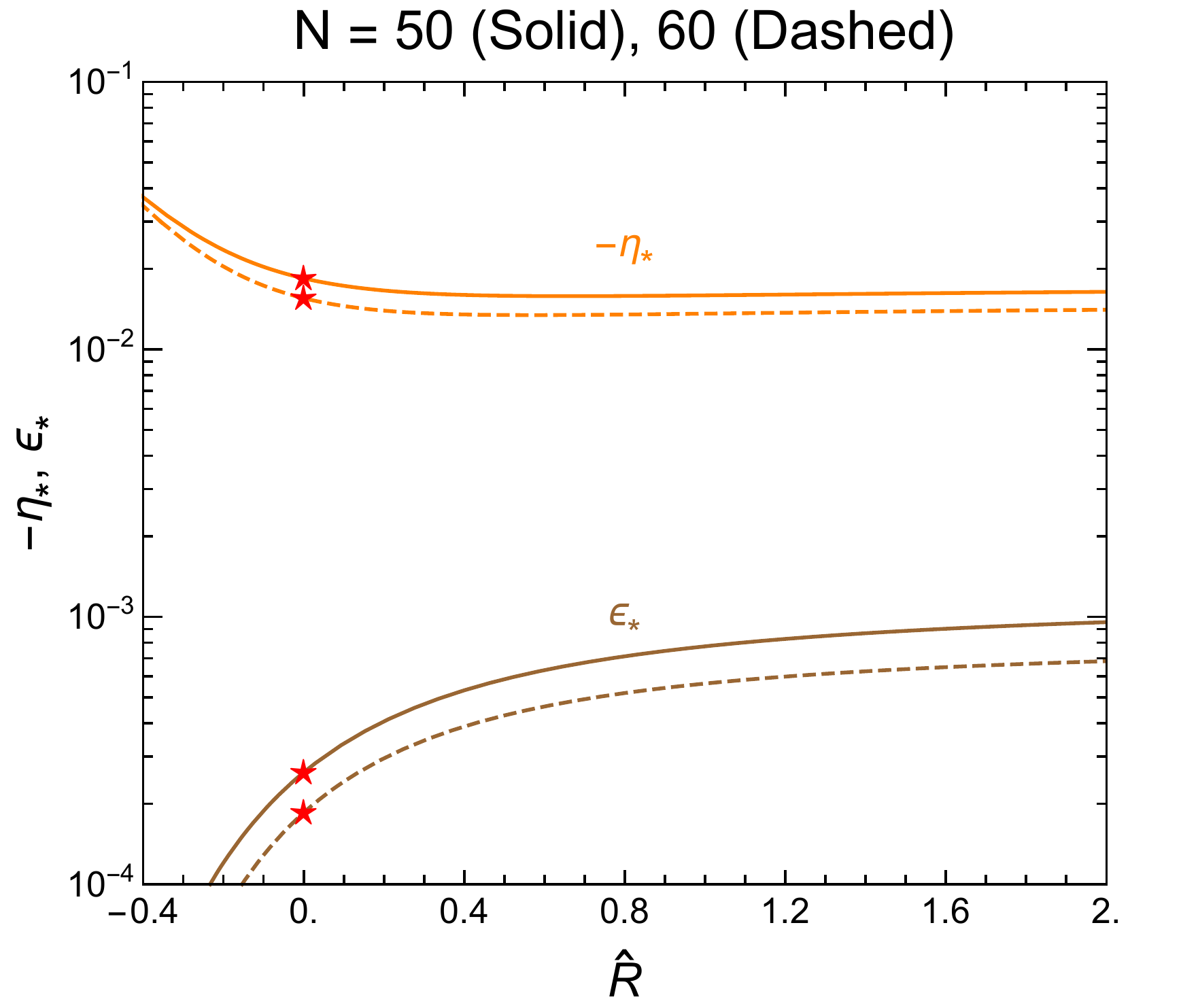}
     \includegraphics[height=0.42\textwidth]{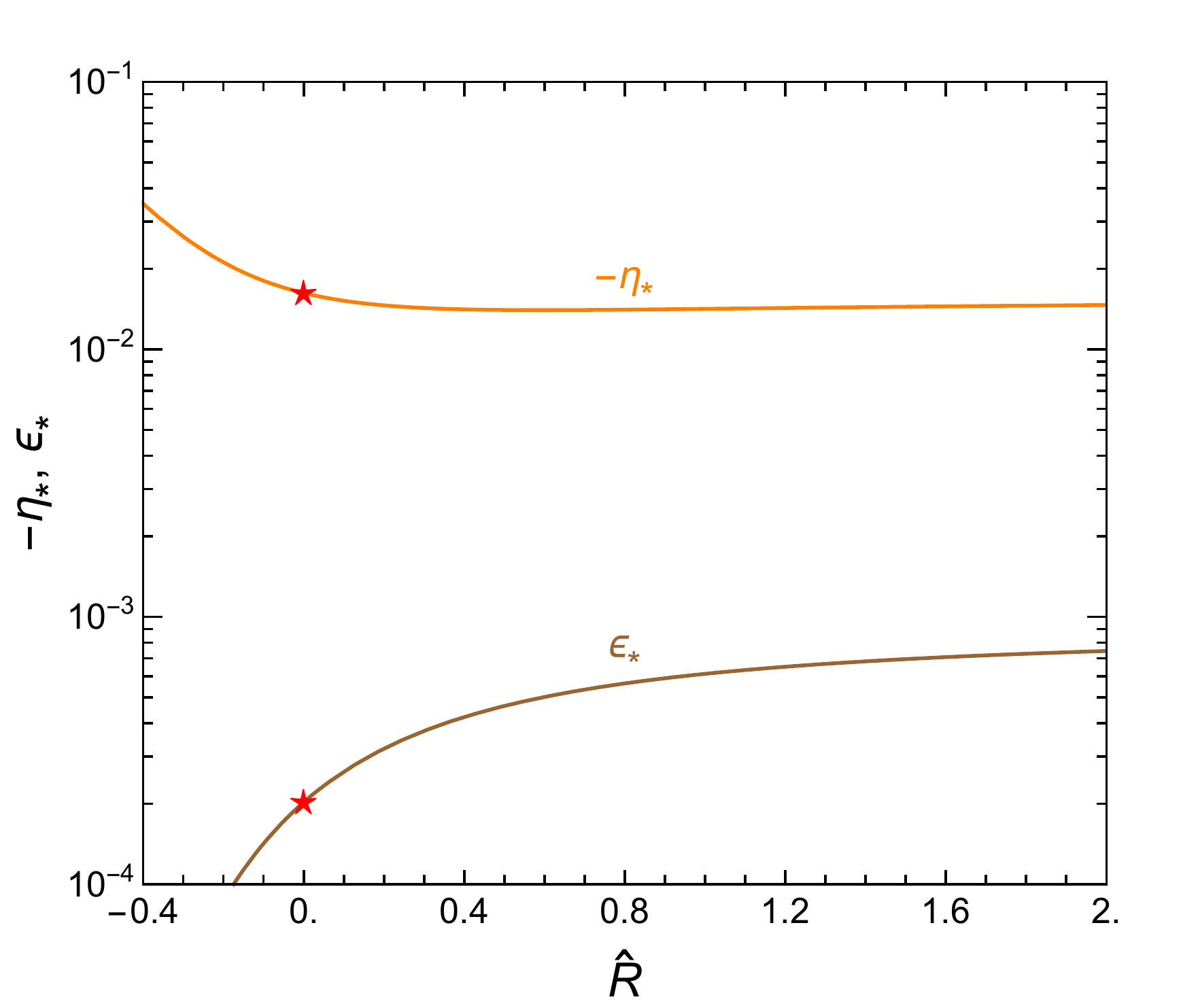}
  \end{center}
  \caption{Left: Slow-roll parameters as a function of ${\hat R}\equiv \xi_1/(\xi_2+\xi_H \tau^2)^{1/2}$. We have chosen the number of efoldings to $N=50, 60$ in solid and dashed lines, respectively. Star points correspond to the case with ${\hat R}=0$. Right: The same as the left plot, but for the case with perturbative reheating. We used $N=55.4+\frac{1}{4}\ln(r/0.01)$.  }
  \label{slowroll}
\end{figure}

Moreover, with eq.~(\ref{epsilon}), the number of efoldings required to solve the horizon problem can be calculated as follows,
\bea
N&=&\int^{\chi_i}_{\chi_f} \frac{{\rm sign}(V'_E)d\chi}{\sqrt{2\epsilon}} \nonumber \\
&\approx &3  A(\chi) \Big(A(\chi)- 
      2 \ln\Big[2 +  {\hat R} \, e^{\frac{1}{\sqrt{6}}\chi} + A(\chi)\Big] \Big) / ({\hat R}^2   
           A(\chi)) \bigg|_{\chi_f}^{\chi_i} \label{Nefold}
\eea
with
\bea
A(\chi)= \sqrt{4  + 12 {\hat R}^2 - 4 {\hat R}^4+ 4 {\hat R}\, e^{\frac{1}{\sqrt{6}}\chi} + 
         {\hat R}^2\, e^{\sqrt{\frac{2}{3}} \chi} }
\eea
where
$\chi_{i,f}$ are the inflaton values at the beginning and end of inflation and we can take $\chi_i=\chi_*$.   Then, we can solve eq.~(\ref{Nefold}) for $\chi_*$ to express the slow-roll parameters at horizon exit in terms of the number of efoldings $N$ and ${\hat R}$. 

\begin{figure}
  \begin{center}
    \includegraphics[height=0.43\textwidth]{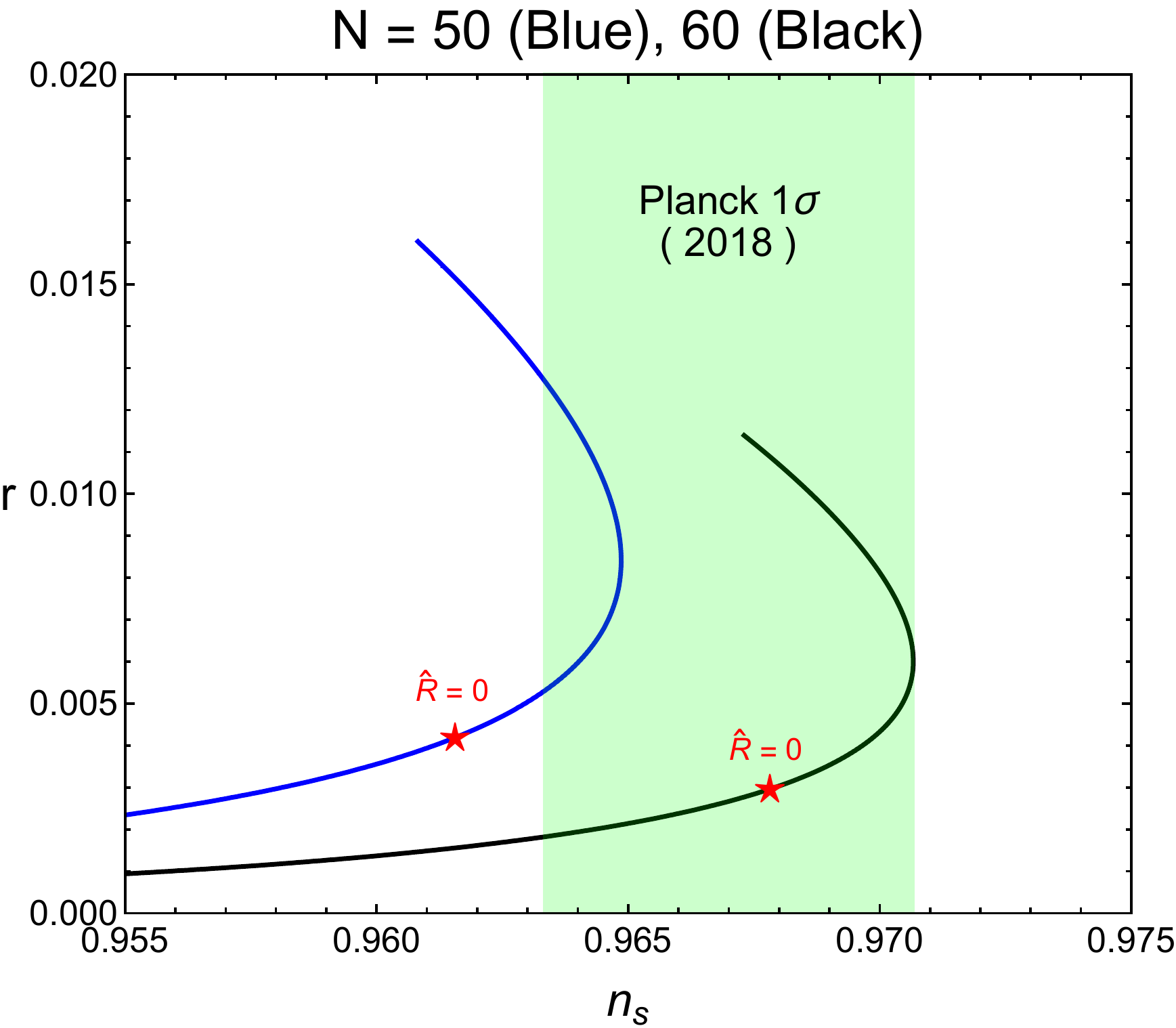} 
        \includegraphics[height=0.43\textwidth]{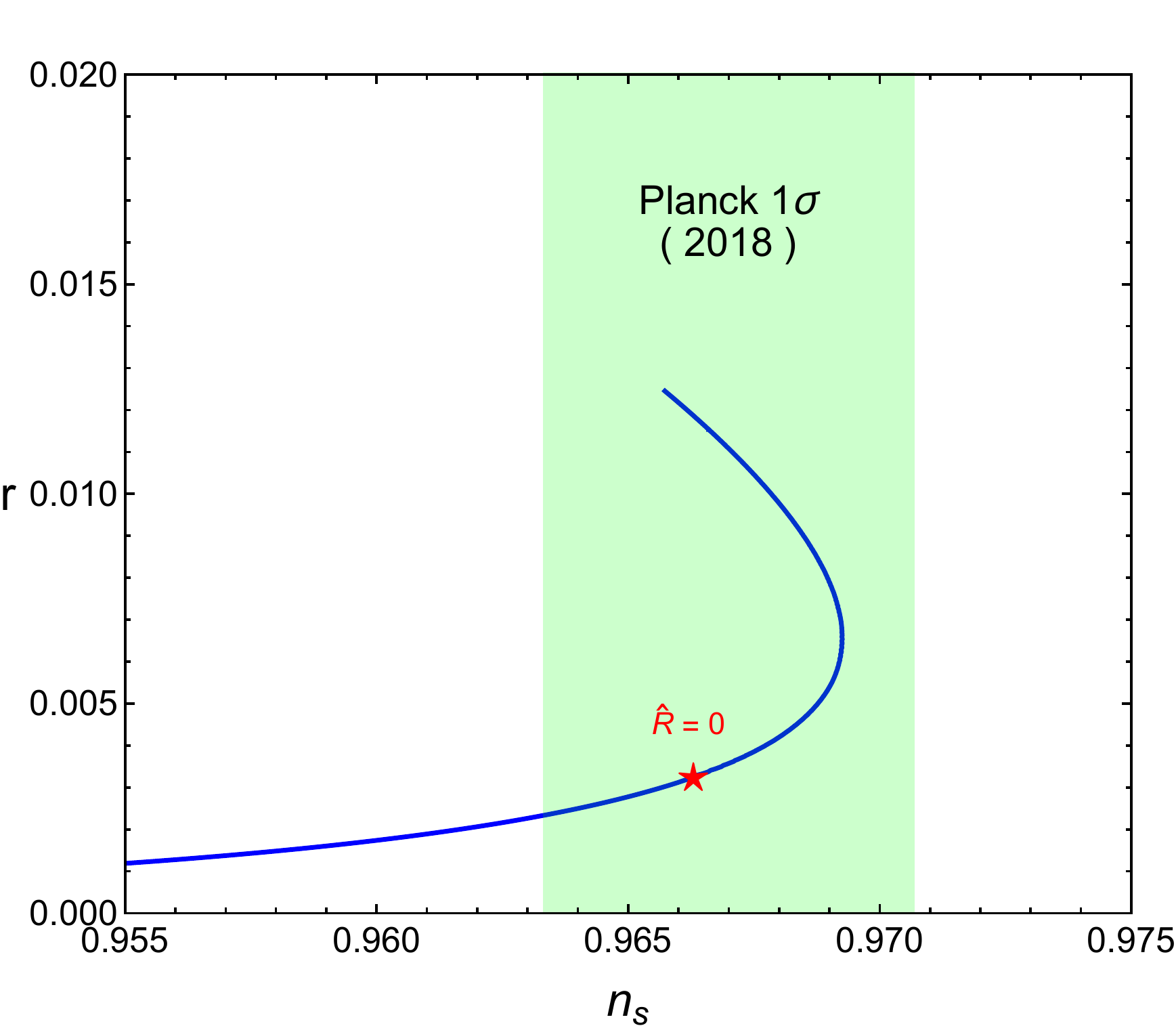} 
  \end{center}
  \caption{Left: Spectral index $n_s$ vs tensor-to-scalar ratio $r$ between ${\hat R}=-0.5$ and $2$. We have chosen $N=50, 60$ in blue and black lines, respectively. Planck $1\sigma$ band is shown in green. Star points correspond to the case with ${\hat R}=0$. Right: The same as the left plot, but for the case with perturbative reheating. We used $N=55.4+\frac{1}{4}\ln(r/0.01)$.}
  \label{ns-r}
\end{figure}

In Fig.~\ref{slowroll}, we show the slow-roll parameters as a function of ${\hat R}$ for $N=50$ and $60$ in solid and dashed lines, respectively, on left, and for $N=55.4+\frac{1}{4}\ln(r/0.01)$ in the case with perturbative reheating, on right. The $\epsilon$ parameter is sensitive to the value of ${\hat R} $, increasing as ${\hat R}$ gets larger. Moreover, in Fig.~\ref{ns-r}, we depict the predictions for the spectral index $n_s$ and the tensor-to-scalar ratio $r$ in our model for $N=50$ and $60$ in blue and black lines, respectively, on left, and for $N=55.4+\frac{1}{4}\ln(r/0.01)$ in the case with perturbative reheating, on right. The values of ${\hat R} $ are taken between $-0.5$ and $2$. In the presence of a sizable ${\hat R}$, the region compatible with Planck 2018 data is enlarged  and the tensor-to-scalar ratio is as large as $0.014$, which is at the detectable level in future CMB experiments.

For ${\hat R}\ll e^{-\frac{1}{\sqrt{6}}\chi_*} $, i.e. $\xi_1\ll 1$, the results with quadratic non-minimal couplings only are recovered, namely, $N\approx\frac{3}{4}\,  e^{\frac{2}{\sqrt{6}}\chi_*}$, $\epsilon_*\approx \frac{4}{3}\,  e^{-\frac{4}{\sqrt{6}}\chi_*} $ and $\eta_*\approx -\frac{4}{3}\,  e^{-\frac{2}{\sqrt{6}}\chi_*}$. Then, we get $\epsilon_*\approx \frac{3}{4N^2}$ and $\eta_*\approx -\frac{1}{N}$, so the spectral index and the tensor-to-scalar ratio become $n_s\approx 1-\frac{2}{N}$ and $r\approx \frac{12}{N^2}$, respectively \cite{Higgsinf}. 

Finally, from the normalization for CMB anisotropies, the vacuum energy during inflation is constrained by $A_s= \frac{1}{24\pi^2} \frac{V_I}{\epsilon_*}\simeq 2.1\times 10^{-9}$ at the Planck pivot scale of $k=0.05\,{\rm Mpc}^{-1}$ \cite{planck2018}. Then, for $\xi_2\gg 1$ and $\xi_H={\cal O}(1)$, depending on the viable inflation vacua given in (\ref{taumin-aa}), we need the combination of quartic couplings and the quadratic non-minimal coupling to satisfy the following conditions, 
\bea
 (1):&&~\frac{1}{\xi_2^2}\Big(\lambda_\sigma -\frac{\lambda^2_{\sigma H}}{4\lambda_H} \Big)=1.2\times 10^{-9}\Big(\frac{r}{0.01} \Big),  \nonumber \\
(2),\,  (4):&& ~\frac{\lambda_\sigma}{\xi_2^2}=1.2\times 10^{-9}\Big(\frac{r}{0.01} \Big).
\label{cobe}
\eea
Therefore, in the case with sigma-Higgs mixed inflation (1), there is a cancellation between quartic couplings, so the CMB normalization can be satisfied for a relatively smaller $\xi_2$. On the other hand,  in the case with sigma inflation (2), (4), the non-minimal coupling $\xi_2$ must be very large unless $\lambda_\sigma$ is small.

\section{Reheating}

In order to discuss the reheating process, we need to identify the form of the inflaton potential at the onset of inflaton oscillation near the minimum. 
We show that the sigma-field potential becomes quartic already at the end of inflation due to a sizable linear non-minimal coupling, so the reheating dynamics becomes different from the usual inflation without a linear non-minimal coupling. We also obtain the reheating temperature of our model, depending on the mixing quartic coupling between sigma and Higgs fields.
Then, we discuss the relevance of preheating and the instability regions for quartic couplings in our model.

\subsection{Inflaton potential during reheating}

For simplicity, we focus on the reheating process of the pure sigma-field inflation but a similar discussion applies for the mixed inflation. 
With $\langle\phi\rangle=v$ during inflation, from eq.~(\ref{sigma-ein}) or (\ref{kinterm}), the canonical inflaton field $\chi$ is related to the original sigma field by
\bea
\frac{d\chi}{d\sigma}&=& \sqrt{\frac{1}{\Omega}+\frac{3\Omega^{\prime 2}}{2\Omega^2}}=\frac{1}{\Omega}\sqrt{1+\xi_1\sigma+\xi_2\sigma^2 +\frac{3}{2}(\xi_1+2\xi_2\sigma)^2} \nonumber \\
&=& \frac{1}{\Omega}\sqrt{1-\frac{\xi^2_1}{4\xi_2}+\xi_2(1+6\xi_2){\bar\sigma}^2 } \label{canonical}
\eea
with ${\bar\sigma}=\sigma+\frac{\xi_1}{2\xi_2}$. 
Therefore, as far as $1 \lesssim \xi_1< 2\sqrt{\xi_2}$, we always get $\bar\sigma\gtrsim \frac{1}{\xi_2}$, so the above equation becomes simplified, independent of $\sigma$ field values, to
\bea
\frac{d\chi}{d\sigma}\approx \sqrt{\frac{3}{2}}\frac{\Omega'}{\Omega}. 
\eea
Thus, we obtain the canonical inflaton field as
\bea
\chi\approx  \sqrt{\frac{3}{2}}\ln\Omega= \sqrt{\frac{3}{2}} \ln (1+\xi_1\sigma+\xi_2\sigma^2) \label{chigood}
\eea 
or
\bea
{\bar\sigma}^2\approx\frac{1}{\xi_2} \left(e^{\frac{2}{\sqrt{6}}\chi}-\Big(1-\frac{R^2}{4} \Big) \right)
\eea
with $R=\xi_1/\sqrt{\xi_2}$. 

As a result, we obtain the inflaton potential in Einstein frame, as follows,
\bea
V_E&=&\frac{\lambda_\sigma}{4\Omega^2}\, \sigma^4 \nonumber \\
&\approx & \frac{\lambda_\sigma}{4\xi^2_2} \left[\Big(1- \Big(1-\frac{R^2}{4}\Big) e^{-\frac{2}{\sqrt{6}}\chi} \Big)^{1/2}-\frac{R}{2}\, e^{-\frac{1}{\sqrt{6}}\chi} \right]^4. \label{potgood}
\eea
Then, we can recover the approximate inflaton potential (\ref{inflapot}) with $\tau=0$ during inflation for ${\bar\sigma}\gtrsim \frac{1}{\sqrt{\xi_2}}$ or $\chi\gtrsim 1$. On the other hand, after the end of inflation, i.e. $\chi\lesssim 1$, we can expand eq.~(\ref{potgood}) to get the inflaton potential during reheating as follows,
\bea
V_E\approx \frac{\lambda_\sigma}{9\xi_1^4}\, \chi^4.  \label{quartic}
\eea
Consequently, we find that the inflaton potential becomes quartic during reheating, due to a sizable linear non-minimal coupling. In this case, the effective quartic coupling for the inflaton becomes suppressed by $\xi^4_1$. The suppressed quartic coupling is due to the redefined sigma field with $\chi\approx\sqrt{\frac{3}{2}}\xi_1 \sigma$ near the true vacuum, as will be discussed in more detail in the next section. 

For comparison, when $\xi_1=0$ or $R=0$ as in Higgs inflation \cite{quadratic},  the inflaton potential (\ref{potgood}) becomes  
\bea
V_E\approx  \frac{\lambda_\sigma}{4\xi^2_2} \left(1- e^{-\frac{2}{\sqrt{6}}\chi} \right)^2.
\eea
The above potential is valid for $\sigma\gtrsim \frac{1}{\xi_2}$, so the inflaton potential becomes quadratic as $V_E\approx  \frac{\lambda_\sigma}{6\xi^2_2}\, \chi^2$ during reheating \cite{quadratic}, unlike the case with $\xi_1\gtrsim 1$. 
Moreover, from eq.~(\ref{canonical}),  when $\xi_1=0$, eq.~(\ref{canonical}) with $\sigma\lesssim \frac{1}{\xi_2}$ leads to $\chi\approx \sigma$, thus the sigma-field potential becomes the same as the one in Jordan frame as $V_E\approx \frac{1}{4}\lambda_\sigma \chi^4$, without a suppression of the quartic coupling.

\subsection{Decays of inflaton condensate and reheating temperature}

In the presence of the quartic inflaton potential during reheating, the inflaton background field (or condensate) $\chi_c$ evolves in time \cite{preheating,quarticpot}, as follows,
\bea
\chi_c(t)=\chi_0(t) \, {\rm cn}  \Big(\omega(t)\, t,\frac{1}{\sqrt{2}} \Big). \label{condensate}
\eea
Here, ${\rm cn}  \Big(\omega(t)\, t,\frac{1}{\sqrt{2}} \Big)\approx \cos(0.85\omega(t))$ with $\omega(t)=2\lambda^{1/2}_\chi \chi_0(t)$ being the oscillation frequency of the inflaton with $\omega(t)\gg H$, and the amplitude of oscillation is given by $\chi_0=\chi_{\rm end}\sqrt{t_{\rm end}/t}$ with $\chi_{\rm end}=(12/\lambda_\chi)^{1/4}\sqrt{H_{\rm end} M_P}$. 
We note that ${\rm cn}(u,m)=\cos \varphi$ is the Jacobi cosine for $u=\int^\varphi_0 d\theta/\sqrt{1-m\sin^2\theta}$. 

From eq.~(\ref{sint}), we consider the relevant Lagrangian for reheating, composed of the inflaton quartic potential and the inflaton interactions in Einstein frame, as follows,
\bea
{\cal L}_{\rm RH}= -\frac{1}{4}\lambda_\chi \chi^4 - \frac{1}{4} \lambda_{\chi H}\, \chi^2 h^2+{\cal L}_{{\rm int},\Omega} \label{RH}
\eea
where $\lambda_\chi\equiv 4\lambda_\sigma/(9\xi^4_1)$ and $ \lambda_{\chi H}\equiv 2\lambda_{\sigma H}/(3\xi^2_1)$ is used, and ${\cal L}_{{\rm int},\Omega}$ is the gravitational inflaton interaction from eq.~(\ref{sint})  during reheating, due to the frame function with eq.~(\ref{chigood}), given by
\bea
{\cal L}_{{\rm int}, \Omega}&=& \frac{2\chi}{\sqrt{6}M_P}\,\left[-\frac{1}{2}(\partial_\mu \phi)^2+2V+\frac{1}{2}m_{f,0}\,\frac{\phi}{v}\,{\bar f}f -\frac{1}{2}\, \delta_V\, m^2_{V,0}\,\frac{\phi^2}{v^2}\, V_\mu V^\mu  \right] \nonumber \\
&\approx&  \frac{2\chi}{\sqrt{6}M_P}\,\left[ -\frac{1}{2}(\partial_\mu h)^2 +m^2_{h,0} h^2 + \frac{1}{2}m_{f,0}{\bar f}f-\frac{1}{2}\, \delta_V\, m^2_{V,0}\,V_\mu V^\mu  \right] +\cdots
\eea

Then, from the quartic terms in the potential in eq.~(\ref{RH}), the inflaton and Higgs boson particles have background-dependent masses due to the inflaton condensate as
\bea
m^2_\chi(t)&=& 3\lambda_\chi \chi^2_c(t) + m^2_{\chi,0}\, ,  \label{chi-mass} \\
m^2_h(t)&=& \frac{1}{2} \lambda_{\chi H}\chi^2_c(t) + m^2_{h,0} \label{h-mass}
\eea
where $m^2_{\chi,0}, m^2_{h,0}$ are inflaton-independent scalar masses. 
As a result, the decay width of the inflaton condensate \cite{quarticpot} is determined to be
\bea
\Gamma_{\chi_c}=\Gamma_{\chi_c\rightarrow \chi\chi}+ \Gamma_{\chi_c\rightarrow hh} \label{infdecay}
\eea
with
\bea
\Gamma_{\chi_c\rightarrow \chi\chi}&=& 0.023 \lambda^{3/2}_\chi \chi_0,  \label{infcc} \\
\Gamma_{\chi_c\rightarrow hh}&=& 0.002 \lambda^2_{\chi H}\lambda^{-1/2}_\chi \chi_0. \label{infhh}
\eea
We note that the gravitational contributions to $\Gamma_{\chi_c\rightarrow hh}$ and additional decay modes can be ignored as far as $\lambda_{\chi H}/\lambda_\chi\gtrsim \frac{3}{4\sqrt{6}}\,\chi_{\rm end}/M_P$ or $\lambda_{\chi H}/\lambda^{3/4}_\chi \gtrsim 1.8\times 10^{-3} (r/0.01)^{1/4}$. 
Henceforth, we assume that this is the case, as will be shown in the later section.

Then, from the condition for the inflaton decoupling at $t_{\rm dec}$, for which
\bea
\Gamma_{\chi_c}=\Gamma_{\chi_c\rightarrow hh}\cdot\Big(\frac{1}{1-{\rm BR}} \Big)\simeq H_{\rm dec}=\sqrt{\frac{\lambda_\chi}{12}}\, \frac{\chi^2_0(t_{\rm dec})}{M_P} \label{decouple}
\eea
with
\be
{\rm BR}=\frac{\Gamma_{\chi_c\rightarrow \chi\chi}}{\Gamma_{\chi_c\rightarrow \chi\chi}+\Gamma_{\chi_c\rightarrow hh}},
\label{BRchi}
\ee 
we obtain the amplitude of the inflaton condenstate as 
\be
\chi_0(t_{\rm dec})=0.007 \lambda^2_{\chi H}\lambda^{-1}_\chi \Big(\frac{1}{1-{\rm BR}} \Big)\, M_P.  \label{dec}
\ee 
Therefore, under the condition of instantaneous reheating,
\bea
\frac{\pi^2 g_*(T_{\rm RH})}{30}\, T^4_{\rm RH}= (1-{\rm BR})\cdot \rho_{\chi_c}(t_{\rm dec})=  (1-{\rm BR})\cdot \frac{\lambda_\chi}{4}\, \chi^4_0(t_{\rm dec}),
\eea
with eq.~(\ref{dec}), we get the reheating temperature as
\bea
T_{\rm RH}&=&0.002\,
\left(\frac{100}{g_*(T_{\rm RH})}\right)^{1/4}  \lambda^2_{\chi H} \lambda_\chi^{-3/4}\, (1-{\rm BR})^{-3/4}\, M_P \nonumber \\
&=&(4.4\times 10^6\,{\rm GeV})\, \left(\frac{100}{g_*(T_{\rm RH})}\right)^{1/4} \left(\frac{\lambda_{\chi H}}{10^{-8}} \right)^2 R^3\, (1-{\rm BR})^{-3/4}\,\Big(\frac{r}{0.01} \Big)^{-3/4} \label{TRH}
\eea
with 
\bea
{\rm BR}=\frac{11.5 \lambda^2_\chi}{11.5\lambda^2_\chi+\lambda^2_{\chi H}}=\frac{0.032R^{-8}\Big(\frac{r}{0.01}\Big)^2}{0.032R^{-8}\Big(\frac{r}{0.01}\Big)^2+\Big(\frac{\lambda_{\chi H}}{10^{-8}}\Big)^2}.
\eea
Here, we used $\lambda_\chi=5.3\times 10^{-10} R^{-4}(r/0.01)$ from the CMB normalization  in eq.~(\ref{cobe}).
Therefore, for $R={\cal O}(1)$ and $r=0.01$, choosing $\lambda_{\chi H}\sim 10^{-8}$, we get ${\rm BR}\ll 1$ and $T_{\rm RH}\sim 10^6\,{\rm GeV}$.

\subsection{Preheating from Higgs portal coupling}

Preheating is a non-perturbative process for reheating and it becomes sometimes dominant.
In our case, since the effective mass of Higgs boson depends on the time-dependent inflaton condensate, this leads to the non-adiabatic excitation of the Higgs perturbation by parametric resonance \cite{preheating0,preheating,review}.  

As discussed in the previous subsection, the inflaton potential becomes quartic during reheating, so the inflaton condensate follows eq.~(\ref{condensate}).  Then, the Fourier mode $h_k$ of the Higgs perturbation with comoving momentum $k$ satisfies the following modified Klein-Gordon equation \cite{preheating},
\bea
{\ddot h}_k + 3 H{\dot h}_k + \Big(\frac{k^2}{a^2}+m^2_h(t) \Big)h_k=0. \label{Hpert}
\eea
Then, redefining the Higgs perturbation by $H_k(t)=a(t) h_k(t)$ and introducing the conformal time by $\eta=\int dt/a(t)$, we can write eq.~(\ref{Hpert}) \cite{preheating} as the Lam\'e equation, 
\bea
H^{\prime\prime}_k+\left(\kappa^2 +\frac{\lambda_{\chi H}}{2\lambda_\chi}\, {\rm cn}^2\Big(x,\frac{1}{\sqrt{2}}\Big) \right) H_k=0 \label{Higgspert}
\eea
where the prime denotes the derivative with respect to the conformal time $\eta$, and $x\equiv \omega(t)t=(48\lambda_\chi)^{1/4}\sqrt{t}$, and the comoving momentum $k$ in units of the initial effective mass of the inflaton is given by
\be
\kappa^2\equiv \frac{k^2}{\lambda_\chi \chi^2_0 \, a(t)^2}.
\ee 
Here, we ignored the inflation-independent Higgs mass, $m^2_{h,0}$, in eq.~(\ref{h-mass}).
Then, the number of Higgs particles created during preheating grows exponentially as $n_k\sim |H_k|^2\sim e^{2\mu_k x}$  with a Floquet index $\mu_k>0$.  When $\frac{{\dot n}_k}{n_k}\sim 2\mu_k {\dot x}\gtrsim \Gamma_h$,  with $\Gamma_h$ being the Higgs decay rate, preheating works for Higgs production.  From $\Gamma_h\sim \frac{y^2_b}{16\pi}\, \langle m_h\rangle$ with $y_b$ being the bottom Yukawa coupling, 
the condition for preheating to work for Higgs production is
\bea
\mu_k\gtrsim 8.3\times 10^{-5} \left(\frac{\lambda_{\chi H}}{10^{-7}} \right)^{1/2} \left(\frac{10^{-10}}{\lambda_\chi} \right)^{1/2}. \label{preheating}
\eea
Here, we note that both $\mu_k {\dot x}$ and $m_h$ are proportional to $\chi_0$, unlike the case with the quadratic inflaton potential where $\mu_k {\dot x}$ is replaced by the inflaton mass, so preheating rate exceeds the Higgs decay rate only if eq.~(\ref{preheating}) is fulfilled.  
On the other hand, if $\mu_k<0$, i.e. outside the instability bands, preheating can be ignored. 

Furthermore, preheating can be dominant over perturbative reheating, provided that $\frac{{\dot n}_k}{n_k}\sim 2\mu_k {\dot x}\gtrsim \Gamma_{\chi_c}\approx  \Gamma_{\chi_c\rightarrow hh}$ with eq.~(\ref{infdecay}),  that is,
\bea
\mu_k\gtrsim 2\times 10^{-7} \left(\frac{\lambda_{\chi H}}{10^{-7}} \right)^2 \left(\frac{10^{-10}}{\lambda_\chi} \right). \label{dominance}
\eea
Therefore, as far as preheating is efficient according to eq.~(\ref{preheating}), it would become a dominant process for reheating.  We note that from $\mu_k {\dot x}\sim m_\chi$, eq.~(\ref{dominance}) is equivalent to $\mu_k m_\chi\gtrsim \Gamma_{\chi_c}$.
If eq.~(\ref{dominance}) is satisfied, the reheating temperature can be determined approximately by the condition, $2\mu_k {\dot x}\sim 3H$, where $H$ is the Hubble parameter during reheating.
In this case, the resulting reheating temperature can be much larger than the one determined by perturbative decay in the previous subsection.

 \begin{figure}
  \begin{center}
    \includegraphics[height=0.45\textwidth]{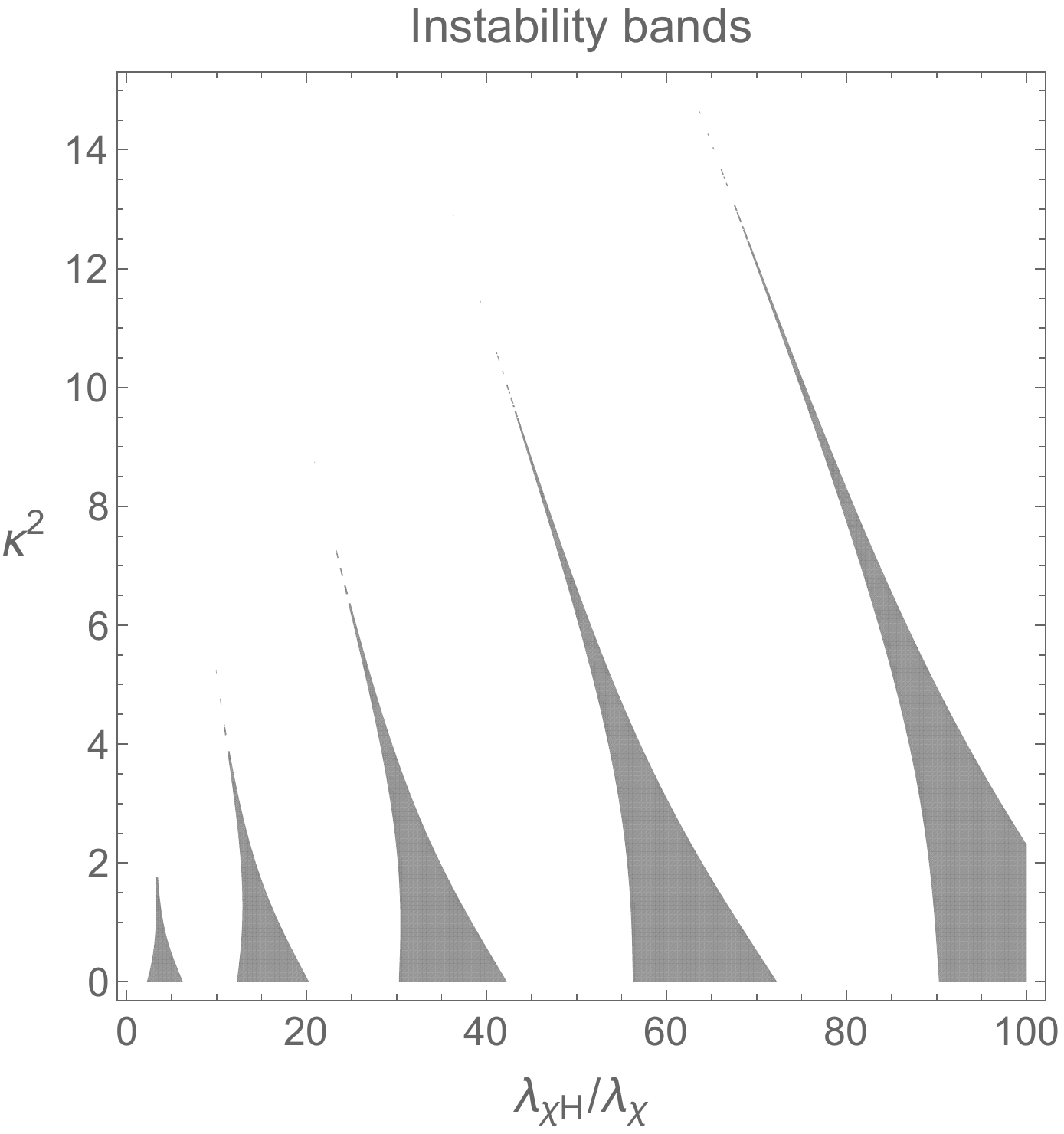} \quad
       \includegraphics[height=0.45\textwidth]{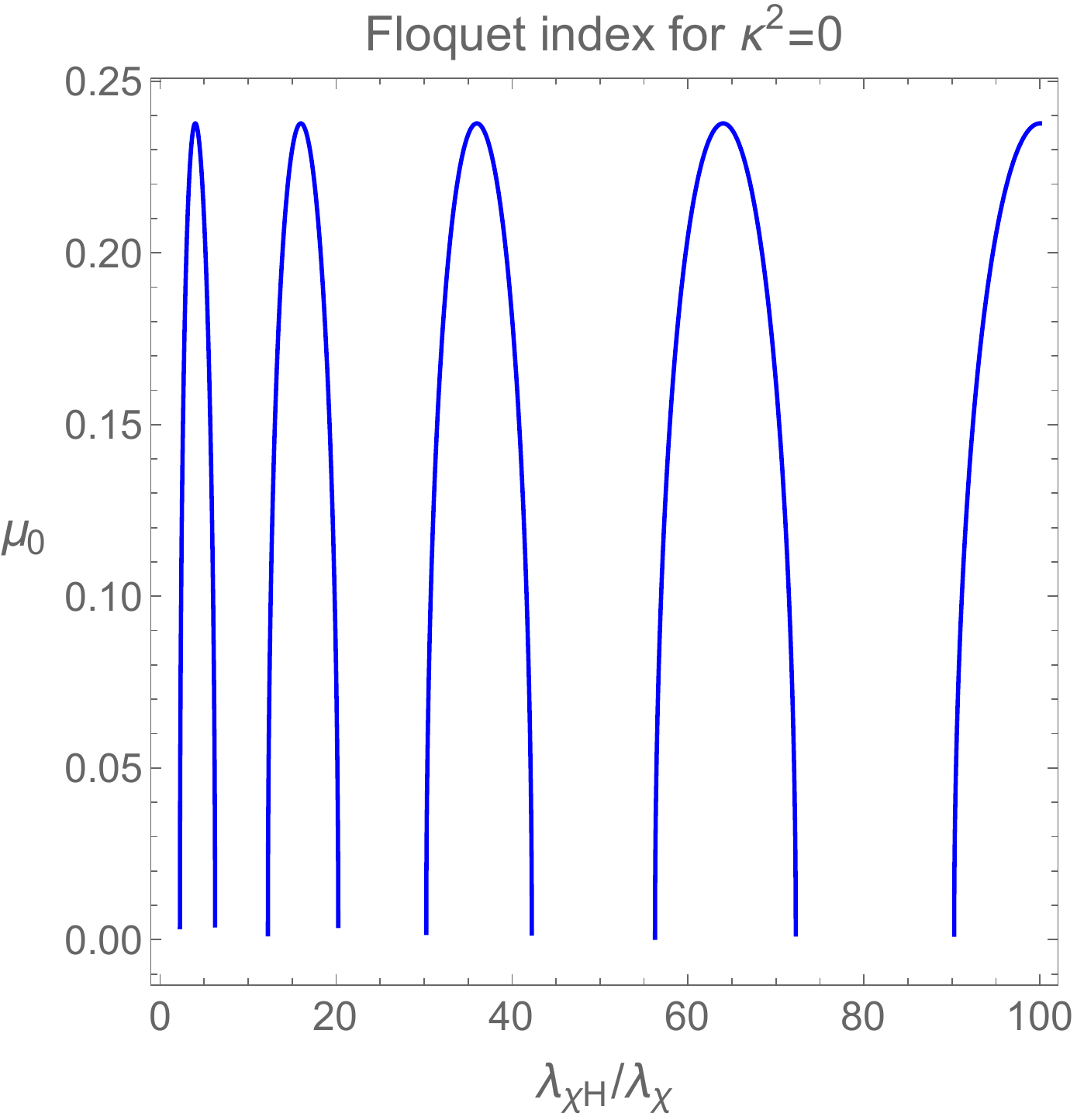} 
  \end{center}
  \caption{(Left) Instability bands for preheating in the parameter space, $\lambda_{\chi H}/\lambda_\chi$ vs $\kappa^2\equiv k^2/(\lambda_\chi \chi^2_0(t)a(t)^2)$.  (Right) Floquet index $\mu_0$ for zero momentum as a function of $\lambda_{\chi H}/\lambda_\chi$. }
  \label{instability}
\end{figure}

In the case of $\lambda_{\chi H}\gtrsim \lambda_\chi$, which is our interest for the later discussion on decaying dark matter, we expand ${\rm cn}\Big(x,\frac{1}{\sqrt{2}}\Big)\approx x$ near $x=0$.
Then, it can be shown that the equation for the Floquet index $\mu_k$ is given \cite{preheating} by
\bea
e^{\mu_k \frac{T}{2}} =|\cos\alpha_k| \sqrt{1+e^{-\pi \gamma^2}} +\sqrt{(1+e^{-\pi \gamma^2})\cos^2\alpha_k-1} \label{floquet}
\eea
where $T=7.416$ is the period of the oscillations in units of $x$, and $\gamma^2\equiv \sqrt{\frac{4\lambda_\chi}{\lambda_{\chi H}}}\, \kappa^2$, and the phase $\alpha_k$ is approximated to
\bea
\alpha_k\approx \pi\sqrt{\frac{\lambda_{\chi H}}{4\lambda_\chi}} +\kappa^2 \sqrt{\frac{\lambda_\chi}{4\lambda_{\chi H}}}\, \ln \frac{\lambda_{\chi H}}{2\lambda_\chi}.  \label{alpha}
\eea
Then, there exists a solution to eq.~(\ref{floquet}), i.e. the exponential growth of created particles is possible, only if
\bea
|\tan\alpha_k|< e^{-\pi\gamma^2/2}.   \label{tanalpha}
\eea

On the left in Fig.~\ref{instability}, we show the instability bands for preheating in the parameter space for $\lambda_{\chi H}/\lambda_\chi$  and the comoving momentum  $\kappa^2\equiv k^2/(\lambda_\chi \chi^2_0(t)a(t)^2)$. On the right in Fig.~\ref{instability}, for zero momentum mode, we draw the Floquet index as a function of $\lambda_{\chi H}/\lambda_\chi$, with its maximum being given by $\mu_{k,{\rm max}}\approx 0.2377$. As a consequence, in the region where eq.~(\ref{tanalpha}) is satisfied, preheating is efficient enough to determine the reheating temperature at a higher value as compared to the case with perturbative decays. 

Preheating process becomes important for broad resonances near the zero momentum mode on the left in Fig.~\ref{instability}. 
However, in the narrow resonances close to cuspy ends of each instability band in the same plot, the redshift of momenta $k$ away from the resonance band can prevent parametric resonance from being efficient \cite{preheating0}. Then, we can estimate the condition for preheating into the Higgs perturbation to be dominant as $\mu_k (\Delta \kappa^2) m_\chi\gtrsim \Gamma_{\chi_c}$ where $ \Delta \kappa^2$ is the width of the narrow resonance. Therefore, the original condition with broad resonances in eq.~(\ref{dominance}) is generalized to $\mu_k\gtrsim  10^{-7} /(\Delta \kappa^2)$. Therefore, for a sufficiently small $\Delta\kappa^2$, we can ignore preheating safely but instead rely on the perturbative decays of the inflaton for reheating, as discussed in the previous subsection.  We assume that this is the case for our later discussion for the calculation of dark matter abundance from the inflaton.

Before ending this subsection, we comment on the inflaton perturbation and preheating.  
The corresponding Lam\'e equation for $X_k(t)=a(t)\delta \chi_k(t)$ with the inflaton perturbation $\delta \chi_k$ is
\bea
X^{\prime\prime}_k+\left(\kappa^2 +3\, {\rm cn}^2\Big(x,\frac{1}{\sqrt{2}}\Big) \right) X_k=0.
\eea
Here, we also ignored the bare inflaton mass, $m^2_{\chi,0}$, in eq.~(\ref{chi-mass}).
In this case, the inflaton perturbation grows for the momenta in the range, $\frac{3}{2}<\kappa^2<\sqrt{3}$ \cite{preheating}.  However, the modes of the inflaton perturbations which are amplified are at sub-Hubble scales during reheating \cite{review}, so there is no effect of the inflaton perturbations produced from preheating at large scales such as CMB \cite{review}. The maximum growth for $\delta \chi_k$ is $\mu_{k,{\rm max}}\approx 0.03598$ at $\kappa^2\approx 1.615$ \cite{preheating}.
If the inflaton perturbation is decoupled from the SM due to small couplings, i.e. $|\lambda_{\chi H}|\lesssim 10^{-7}$, the produced inflaton would not thermalize the SM particles.

\section{Non-minimal couplings and unitarity scales}

We discuss the impacts of the linear non-minimal coupling in identifying the physical parameters of the scalar potential in the vacuum and show how the unitarity problem with a large quadratic non-minimal coupling can be eliminated by an appropriate linear non-minimal coupling.

\subsection{Physical parameters in the vacuum}

Taking $\sigma,h\ll 1$ near the vacuum, we get the approximate quadratic kinetic terms in eq.~(\ref{kinterm}) as 
\bea
{\cal L}_{{\rm kin},0}= \frac{1}{2} \Big(1+\frac{3}{2} \xi^2_1 \Big)(\partial_\mu\sigma)^2+\frac{1}{2}(\partial_\mu \phi)^2.
\eea
Then, from the canonical sigma field,
\bea
\chi= \Big(1+\frac{3}{2} \xi^2_1\Big)^{1/2} \sigma, \label{rescale}
\eea
the frame function becomes
\be
\Omega=1+\frac{\xi_1}{\sqrt{1+\frac{3}{2}\xi^2_1}}\,\chi+\frac{\xi_2}{1+\frac{3}{2}\xi^2_1}\,\chi^2+\xi_H \phi^2.  \label{newframe}
\ee
Moreover, we get the Einstein-frame potential (\ref{einstein-pot}) for the canonical sigma field $\tilde \sigma$, as follows,
\bea
V_E(\sigma,\phi)\approx V= V_0+ \frac{1}{2} m^2_\chi \chi^2 +\frac{1}{4} \lambda_\chi \chi^4 +\frac{1}{4} \lambda_{\chi H} \chi^2  \phi^2 +\frac{1}{2} m^2_H \phi^2  +\frac{1}{4}\lambda_H \phi^4 \label{potil}
\eea
with
\bea
m^2_\chi &=&  \Big(1+\frac{3}{2} \xi^2_1 \Big)^{-1} m^2_\sigma\approx \frac{2}{3} \,\frac{m^2_\sigma}{\xi^2_1}, \\
\lambda_\chi &=& \Big(1+\frac{3}{2} \xi^2_1 \Big)^{-2}  \lambda_\sigma\approx \Big(\frac{2}{3}\Big)^2\, \frac{\lambda_\sigma}{\xi^4_1}, \\
\lambda_{\chi H} &=& \Big(1+\frac{3}{2} \xi^2_1 \Big)^{-1} \lambda_{\sigma H}\approx \frac{2}{3}\, \frac{\lambda_{\sigma H}}{\xi^2_1}. 
\eea
On the other hand, the interaction terms containing $\phi$ only do not rescale. 
Therefore, if dimensionful and dimensionless parameters are of common origin in Jordan frame, we can get a natural hierarchy of masses and couplings for $\xi_1\gg 1$: $|m_\chi| \ll |m_H|$, and $\lambda_\chi, |\lambda_{\chi H}|\ll \lambda_H$. 
 
After electroweak symmetry breaking, the effective mass of the inflaton has a tree-level shift as $m^2_{\chi,{\rm eff}} = m^2_\chi + \frac{1}{2} \lambda_{\chi H} v^2$, due to the mixing Higgs quartic coupling.  Higgs loop corrections to the inflaton mass is $\Delta m^2_\chi\sim\frac{\lambda_{\chi H}}{16\pi^2}\, m^2_H$, so they are subdominant as compared to the tree-level shift. 
The mass shift of the inflaton is much smaller than the Higgs mass for $|\lambda_{\chi H}|\ll 1$,.
In a later discussion for light inflaton dark matter, however, we need to tune the bare inflaton mass $m^2_\chi$ against $\lambda_{\chi H}$ for a phenomenologically desirable mass, such as for the relic density.  For simplicity, henceforth we use the same notation for the effective inflaton mass as $m^2_\chi$.

\subsection{Unitarity scales}

 \begin{figure}
  \begin{center}
    \includegraphics[height=0.50\textwidth]{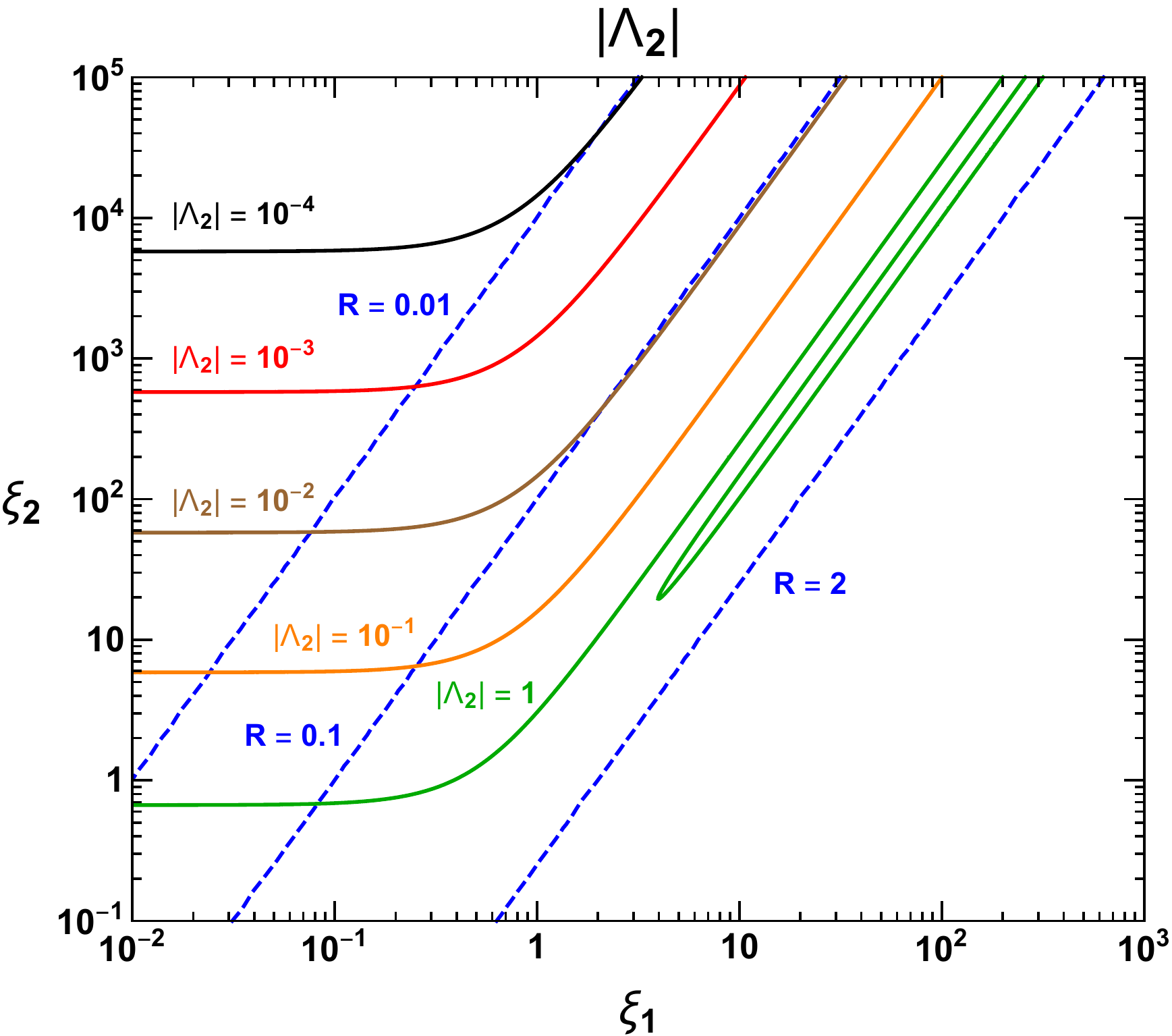}
  \end{center}
  \caption{Contours of $|\Lambda_2|$  in units of $M_P$ in the plane of $\xi_2$ and $\xi_1$ in solid lines.  We overlaid in blue dashed lines the contours of $R=\xi_1/\sqrt{\xi_2}$.}
  \label{cutoff}
\end{figure}

In terms of the canonical sigma field, we obtain the leading derivative interaction terms \cite{hmlee} from eq.~(\ref{kinterm}),
\bea
{\cal L}_{\rm int}&=& -\frac{1}{\Lambda_1}\, \chi(\partial_\mu\chi)^2+ \frac{1}{\Lambda^2_2}\, \chi^2(\partial_\mu\chi)^2 -\frac{1}{\Lambda^2_3}\,\phi^2(\partial_\mu\chi)^2 -\frac{1}{\Lambda_4}\,\chi(\partial_\mu \phi)^2-\frac{1}{\Lambda^2_5}\, \chi^2(\partial_\mu \phi)^2 \nonumber \\
&&  -\frac{1}{\Lambda^2_6}\, \phi^2(\partial_\mu \phi)^2 +\frac{1}{\Lambda_7}\, \phi(\partial_\mu\chi)(\partial^\mu \phi)-\frac{1}{\Lambda^2_8}\, \chi \phi(\partial_\mu\chi)(\partial^\mu \phi) +\cdots
\eea
where the ellipses are higher dimensional terms and the cutoff scales in the leading terms read
\bea
\Lambda_1&\equiv &\frac{2\Big(1+\frac{3}{2}\xi^2_1\Big)^{3/2}}{\xi_1(1+3\xi^2_1-6\xi_2)} \approx \sqrt{\frac{3}{2}}\, \frac{R^2}{R^2-2}\,, \\
|\Lambda_2|&\equiv &\frac{\sqrt{2}\Big(1+\frac{3}{2}\xi^2_1\Big)}{\Big|\xi^2_1\Big(1+\frac{9}{2}\xi^2_1\Big)-\xi_2 (1+15\xi^2_1-6\xi_2)\Big|^{1/2}}\approx \frac{R^2}{\Big|R^4-\frac{2}{3}\Big(5R^2-2 \Big) \Big|^{1/2}}\,, \\
\Lambda_3&\equiv&\sqrt{\frac{2\Big(1+\frac{3}{2}\xi^2_1\Big)}{\xi_H (1+3\xi^2_1)}}\approx \frac{1}{\sqrt{\xi_H}}, \\
\Lambda_4 &=&\frac{2\sqrt{1+\frac{3}{2}\xi^2_1}}{\xi_1}\approx \sqrt{6} \,, \\
|\Lambda_5|&=&\sqrt{ \frac{2\Big(1+\frac{3}{2}\xi^2_1\Big)}{|\xi_2-\xi^2_1|}}\approx  \sqrt{\frac{3 R^2}{|1-R^2|}}, \\
|\Lambda_6|&=&\sqrt{\frac{2}{|\xi_H(1-6\xi_H)|}}, \\
\Lambda_7 &=&\frac{\sqrt{1+\frac{3}{2}\xi^2_1}}{3\xi_H \xi_1}\approx \frac{1}{\sqrt{6}\,\xi_H} , \\
|\Lambda_8|&=&\sqrt{\frac{1+\frac{3}{2}\xi^2_1}{6\xi_H |\xi_2-\xi^2_1|}}\approx \frac{1}{2}\sqrt{\frac{R^2}{\xi_H|1-R^2| }}\,.
\eea
with
\be
R\equiv \frac{\xi_1}{\sqrt{\xi_2}}.
\ee
Here, we assumed $\xi_1\gg 1$ in the above approximations. 
For $\xi_H={\cal O}(1)$ and $\xi_1\gg 1$, the unitarity scales depend only on the ratio of the non-minimal couplings, $R$. 
In Fig.~\ref{cutoff}, as a representative example, we draw the contour plot for the unitarity scale $|\Lambda_2|$ in the parameters space for $\xi_1$ and $\xi_2$, showing that $\xi_1$ is saturated to $\sqrt{\xi_2}$ in order to maintain $|\Lambda_2|$ of order the Planck scale for a large $\xi_2$.

\section{Inflaton couplings to the SM at low energies}

The sigma field also has dilaton-like couplings to the SM through the trace of the energy-momentum tensor, due to the linear non-minimal coupling.
In this section,  we consider all the linear couplings of the sigma field to the SM in the low energy in Einstein frame. 

From $\Omega=1+\xi_1\sigma+ \xi_2 \sigma^2+\xi_H \phi^2$ and eq.~(\ref{potil}), we expand the inflaton interaction Lagrangian (\ref{sint}) in Einstein frame to identify the linear coupling of the canonical sigma field as follows,
\bea
{\cal L}_{{\rm int},\xi_1}&=& \xi_1\sigma\left[-\frac{1}{2}(\partial_\mu \phi)^2+2V+\frac{1}{2}m_{f,0}\,\frac{\phi}{v}\,{\bar f}f -\frac{1}{2}\, \delta_V\, m^2_{V,0}\,\frac{\phi^2}{v^2}\, V_\mu V^\mu  \right]\nonumber \\
&=& \frac{1}{2}  \xi_1\sigma\, T^\mu_{0,\mu} \nonumber \\
&=& \frac{1}{2}  \frac{ \xi_1}{\sqrt{1+\frac{3}{2}\xi^2_1}}\, \frac\chi{M_P}\, T^\mu_{0,\mu}
 \label{sigco}
\eea
where $T^\mu_{0,\mu}$ is the trace of the energy-momentum tensor at tree level on the equations of motion and use is made of the canonical sigma field (\ref{rescale}) in the true vacuum. 
Here, we have recovered the Planck scale $M_P$, and $\xi_1$ is given in units of $M_P$. 

We remark that the minimal couplings of gauge bosons to SM fermions do not depend on the frame function $\Omega$, so there is no coupling between the sigma field and one SM gauge boson. 
Since the covariant derivative terms of fermions do not contribute to the trace of the energy-momentum tensor, our results confirm that minimal couplings between the sigma field and one SM gauge boson are absent, unlike the approach of Ref.~\cite{ibarra} where these couplings however arise at higher orders in perturbation theory.

\subsection{Inflaton couplings to massive particles}

From eq.~(\ref{sigco}), we consider the linear couplings of the sigma field  to the Higgs field,
\bea
{\cal L}_h =\frac{1}{2}  \frac{ \xi_1}{\sqrt{1+\frac{3}{2}\xi^2_1}}\, \frac\chi{M_P}\left[ -(\partial_\mu \phi)^2 + 4 V(\chi, \phi)\right].  \label{Hint}
\eea
The tadpole term for $\chi$ vanishes in the vacuum with a vanishingly small cosmological constant, $\langle V(\chi,\phi)\rangle\approx 0$, leading to an extremely tiny VEV of the sigma field, thus the Higgs-sigma mixing is negligible.  Thus, this is different from the case where a light inflaton carries a sizable Higgs mixing due to a sizable inflaton VEV so it has Higgs-like couplings to the SM \cite{Linflaton}.
Moreover, the mass mixing vanishes in the minimum of the potential $U(\chi,\phi)$. 
We note that the effective mass of the sigma field is shifted to $m^2_\chi+\frac{1}{2} \lambda_{\chi H} v^2$ after electroweak symmetry breaking, but we keep the same notation for the sigma field mass as $m_\chi$ for simplicity.

As a consequence, expanding the Higgs field about the vacuum as $\phi=v+h$, the linear couplings of the sigma-like field (\ref{Hint}) become
\bea
{\cal L}_{h}
&=& \frac{1}{2}\frac{ \xi_1}{\sqrt{1+\frac{3}{2}\xi^2_1}}\, \frac\chi{M_P}\bigg[-(\partial_\mu h)^2+2m^2_\chi\,\chi^2+2m^2_h h^2+\lambda_\chi \chi^4 +4\lambda_H v \, h^3 \nonumber \\
&&+\lambda_H h^4+2\lambda_{\chi H} v\, \chi^2 h+\lambda_{\chi H}\chi^2 h^2 \bigg].
\eea
Then, the sigma field decays into a pair of Higgs bosons, on-shell or off-shell, through the Higgs kinetic term and mass term.  
We note that the Feynman rule for the vertex with one sigma field and two Higgs bosons with outgoing momenta, $p_1$ and $p_2$, is given by
\bea
V_{{\chi hh}}=  \frac{i}{M_P}\frac{ \xi_1}{\sqrt{1+\frac{3}{2}\xi^2_1}}\, \Big(2m^2_h+p_1\cdot p_2 \Big).
\eea

From eq.~(\ref{sigco}), with $\phi=v+h$, we get the linear couplings of the sigma field to massive fermions and electroweak gauge bosons as
\bea
{\cal L}_f &=& \frac{1}{2}\frac{ \xi_1}{\sqrt{1+\frac{3}{2}\xi^2_1}}\,\frac\chi{M_P}\left( m_{f,0} {\bar f} f + \frac{m_{f,0}}{v}\,h  {\bar f} f\right)
\equiv g_{\chi ff}\, \chi  {\bar f} f+\cdots, \\  \label{Yukawa}
{\cal L}_V
&=& -\frac{1}{2} \frac{ \xi_1}{\sqrt{1+\frac{3}{2}\xi^2_1}}\,\frac\chi{M_P} \left(\delta_V\,m^2_{V,0} V_\mu V^\mu+2 \frac{\delta_V\,m^2_{V,0}}{v}\, h\, V_\mu V^\mu+  \frac{\delta_V\,m^2_{V,0}}{v^2}\, h^2\, V_\mu V^\mu\right) \nonumber \\
&\equiv& g_{\chi VV}\, \chi V_\mu V^\mu +\cdots. 
\label{Gauge}
\eea
Then, the sigma field can decay into a pair of SM fermions or gauge bosons.
If the sigma field is lighter than pions, it can decay dominantly into a pair of muons for $m_\chi>2m_\mu$ or a pair of electrons for $m_\chi<2m_\mu$.

\subsection{Inflaton couplings to massless gauge bosons}

We now consider the sigma field couplings to massless gauge bosons. 
In this case, there are two contributions coming from trace anomalies and threshold effects due to heavy particles.

First we note that the trace of the energy-momentum tensor is corrected due to scale anomalies at loop order to the following,
\bea
T^\mu_\mu= T^\mu_{0,\mu}+ \frac{\beta_S(\alpha_S)}{4\alpha_S}\, G^a_{\mu\nu}G^{a\mu\nu} +\frac{\beta_{\rm EM}(\alpha)}{4\alpha}\, F_{\mu\nu}F^{\mu\nu} \label{trace}
\eea
where $\beta_S(\alpha_S)$ and $\beta_{\rm EM}(\alpha)$ are the beta functions for $\alpha_S=\frac{g^2_S}{4\pi}$ and $\alpha=\frac{e^2}{4\pi}$, respectively, and they are given at one loop by $\beta_S=-\frac{\alpha^2_S b_3}{2\pi}$ and $\beta_{\rm EM}=-\frac{\alpha^2 b_\gamma}{2\pi}$ where $b_3, b_\gamma$ are beta function coefficients in the SM, given by $b_3=7$ and $b_\gamma=-\frac{11}{3}$, respectively.  
Therefore, with $T^\mu_{0,\mu}$ in eq.~(\ref{sigco}) being replaced $T^\mu_\mu$, there are sigma field couplings to two photons and two gluons.

Furthermore, the sigma field couples to massive particles through the energy-momentum tensor in $T^\mu_{0,\mu}$. Since all the SM particle have sigma-field dependent masses, $m^2=\Omega^{-1}m^2_0$, where $m^2_0$ being independent of the sigma field, they contribute to the effective QED gauge coupling  at the threshold, $q^2=m^2$, as
\be
 \frac{1}{e^2(m)}= \frac{1}{e^2(\Lambda)}-\frac{B_\gamma}{16\pi^2}\,\ln \Big(\frac{\Lambda^2}{\Omega^{-1} m^2_0}\Big)
\ee
with $\Lambda$ is the cutoff scale and $e(\Lambda)$ is the QED gauge coupling at the cutoff scale.
Therefore, from the gauge kinetic term, $-\frac{1}{4e^2(m)}\, F_{\mu\nu} F^{\mu\nu}$, after absorbing the gauge coupling by the gauge field with $A_\mu\rightarrow  e A_\mu$ , we obtain the additional contributions to the effective sigma field coupling to photons, as follows,
\bea
\Delta {\cal L}_\gamma
&=&  \frac{B_\gamma \, \alpha}{16\pi} \frac{\xi_1}{\sqrt{1+\frac{3}{2}\xi^2_1}}\,  \frac\chi{M_P}\,\, F_{\mu\nu} F^{\mu\nu}
\eea
where the beta function coefficient for EM gauge coupling is given by
\bea
B_\gamma=\sum_f b_f + b_W
\eea
with $b_W=7$ and $b_f=-\frac{4}{3}N_c Q^2_f$. 
Here, the sum $\sum_f$ is performed over all the SM charged fermions heavier than the typical energy scale, for instance, the inflaton mass, in case of the inflaton decay process.  
For example, for $m_\mu\lesssim m_\chi\lesssim m_c$, $B_\gamma=\sum_{f=c,\tau,b,t}b_f+b_W=\frac{5}{3}$; for $m_e \lesssim m_\chi\lesssim m_\mu$, $B_\gamma=-\frac{7}{3}$; for $m_\chi\lesssim m_e $, $B_\gamma=-\frac{11}{3}$.
For  $m_\chi<2m_e$, it decays dominantly into a photon pair.

Consequently, including the sigma field coupling due to trace anomalies in eq.~(\ref{trace}),
we obtain the full  effective sigma field coupling to photons as
\bea
{\cal L}_\gamma &=&  {\cal L}_{\gamma,{\rm trace}} + \Delta {\cal L}_{\gamma} \nonumber \\
&=& - \frac{b_{\gamma,L}\, \alpha}{16\pi} \frac{\xi_1}{\sqrt{1+\frac{3}{2}\xi^2_1}}\,  \frac\chi{M_P}\,\, F_{\mu\nu} F^{\mu\nu}\equiv g_{\chi \gamma\gamma}  \, \chi\,\, F_{\mu\nu} F^{\mu\nu}
\label{photon}
\eea
with $b_{\gamma,L}\equiv b_\gamma-B_\gamma$ being the beta function coefficient for EM gauge coupling due to light charged particles. Thus, the contributions from heavy charged particles cancel the counterpart of scale anomalies, leaving the scale anomalies from light charged particles.

Similarly, the threshold corrections to the running QCD gauge coupling due to heavy quarks lead to the additional contribution to the sigma field couplings to gluons, as follows,
\bea
\Delta{\cal L}_g =  \frac{B_3 \, \alpha_S}{16\pi} \frac{\xi_1}{\sqrt{1+\frac{3}{2}\xi^2_1}}\,  \frac\chi{M_P}\,\, G^a_{\mu\nu} G^{a\mu\nu}
\eea
where $B_3$ is the QCD beta function coefficient due to $n_H$ heavy quarks, given by $B_3=-\frac{2}{3} n_H$. 
Consequently, including the sigma field coupling due to trace anomalies in eq.~(\ref{trace}),
we obtain the full  effective sigma field coupling to photons as
\bea
{\cal L}_g &=&  {\cal L}_{g,{\rm trace}} + \Delta {\cal L}_{g} \nonumber \\
&=&  -\frac{b_{3,L} \, \alpha_S}{16\pi} \frac{\xi_1}{\sqrt{1+\frac{3}{2}\xi^2_1}}\,  \frac\chi{M_P}\,\, G^a_{\mu\nu} G^{a\mu\nu}\equiv g_{\chi gg}  \, \chi\,\, G^a_{\mu\nu} G^{a\mu\nu}
\label{gluon}
\eea
with $b_{3,L}\equiv b_3-B_3$ being the beta function coefficient for strong gauge coupling due to light quarks and gluons. Thus, similarly to the case with photon couplings, there is a cancellation between the contributions from heavy colored quarks and the counterpart of scale anomalies, so the scale anomalies from light quarks and gluons remain.

If the sigma field is heavier than $1.5\,{\rm GeV}$, we can consider the sigma field decays into a gluon pair.  But, for $1.5\,{\rm GeV}<m_\chi<2.5\,{\rm GeV}$, either descriptions in terms of mesons or quarks/gluons are not quite correct \cite{radion1}. For $m_\chi>2.5\,{\rm GeV}$, we can use the description of quarks/gluons and the coefficient of the QCD beta function, which is given by $b_{3,L}=11-\frac{2}{3} n_L$ from $n_L$ light quarks only.
 
We note that the effects of heavy particle masses in the effective inflaton couplings to photons and gluons, $g_{\chi\gamma\gamma}$ and $g_{\chi gg}$, can be taken into account through the loop functions as shown in eqs.~(\ref{decaygaga}) and (\ref{decaygg}) of the appendix.

\subsection{Inflaton couplings to mesons}

When the sigma field is lighter than $1.5\,{\rm GeV}$, we need to include the sigma field decays into a pair of mesons in chiral perturbation theory, instead of quarks or gluons.
In this case, we need to take the beta function coefficient of strong gauge coupling as $b_{3,L}=\frac{29}{3}(9)$ for $m_\chi>2m_\pi(2m_K)$ for $u,d (u,d,s)$ light quarks in chiral perturbation theory \cite{radion}.

 \begin{figure}
  \begin{center}
    \includegraphics[height=0.40\textwidth]{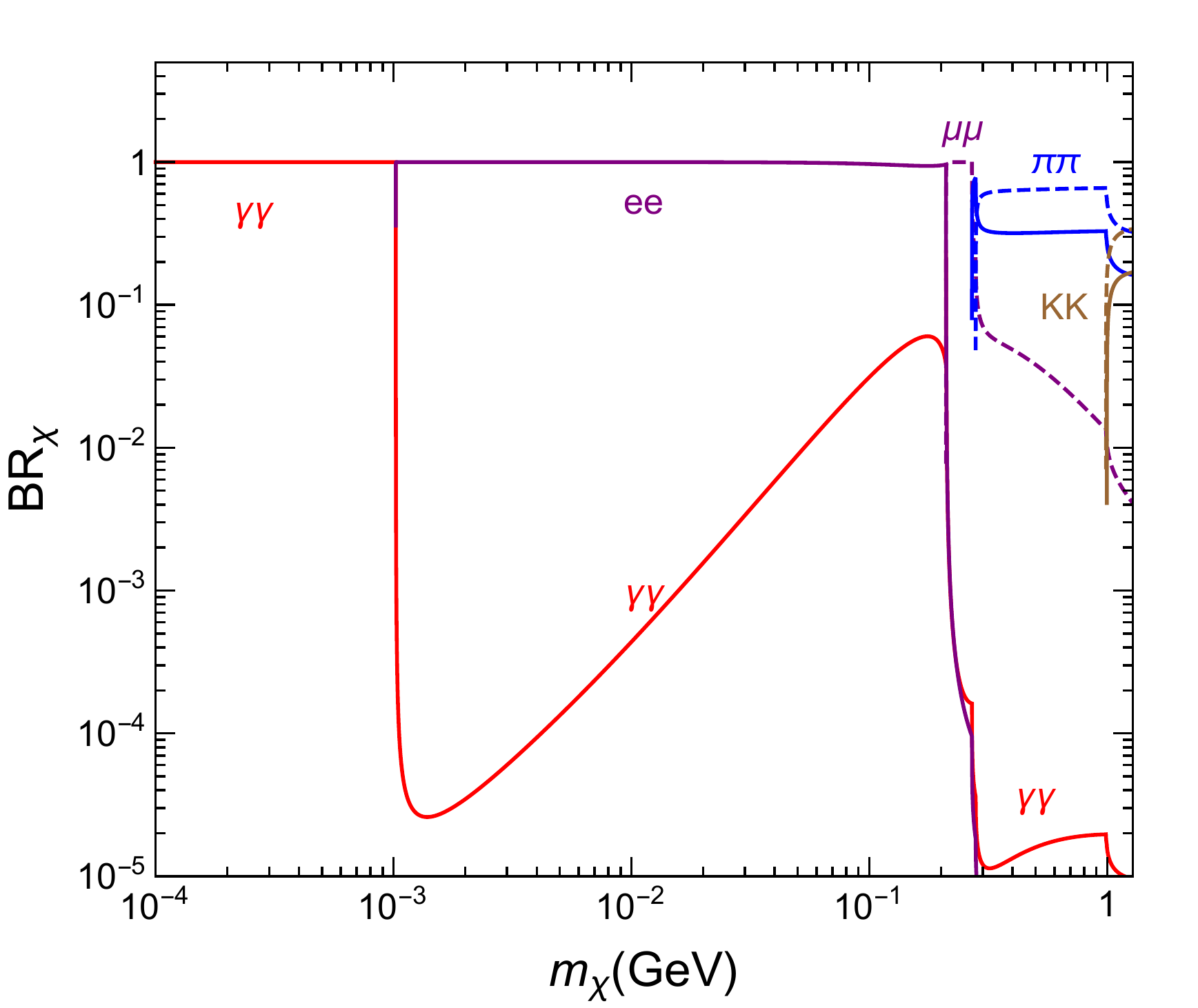}\,\,\,
     \includegraphics[height=0.40\textwidth]{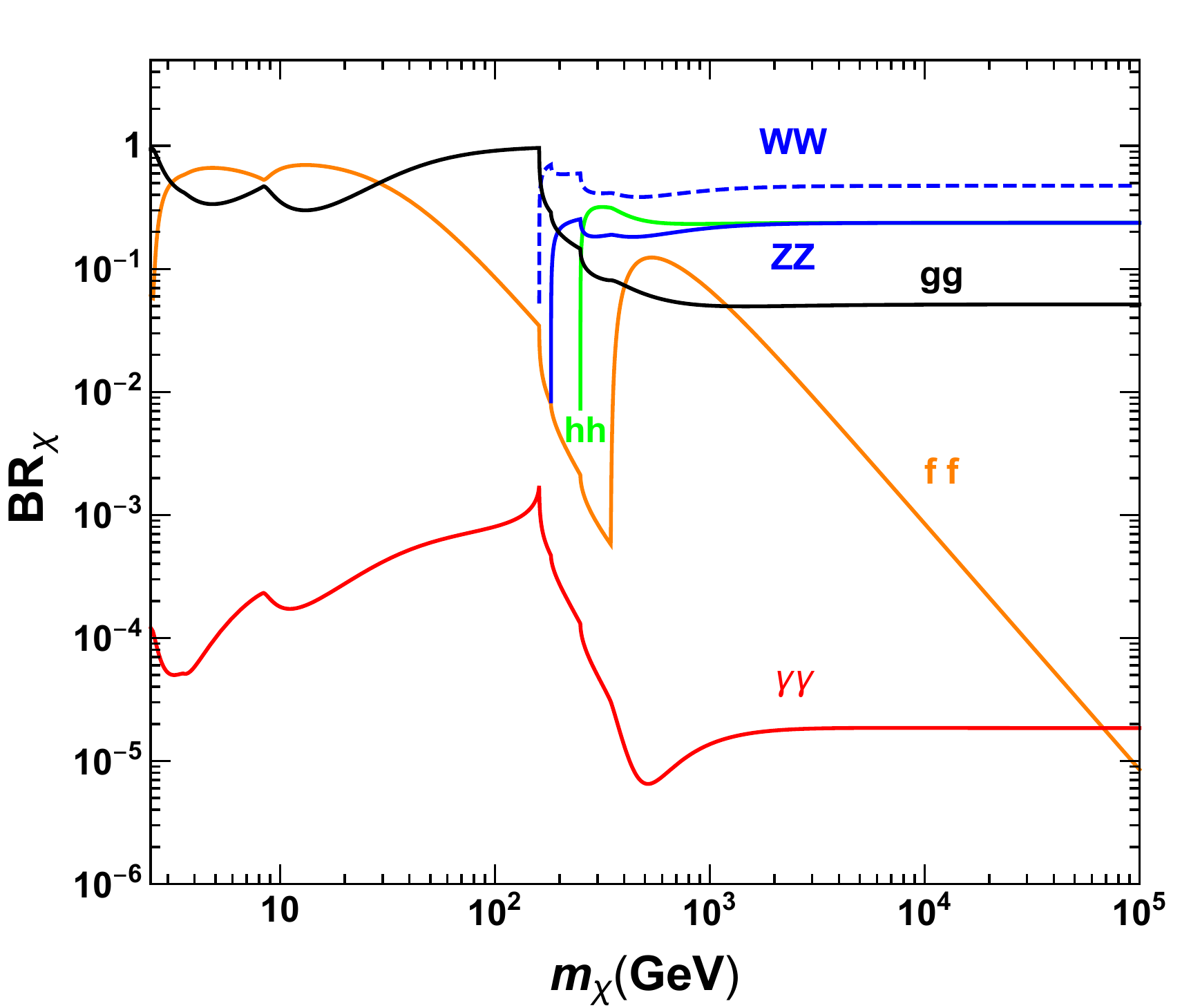}
  \end{center}
  \caption{Decay branching ratios of  the inflaton as a function of $m_\chi$. Inflaton masses are taken to $10^{-4}\,{\rm GeV}<m_\chi<m_c$ on left and $2.5\,{\rm GeV}<m_\chi<10^5\,{\rm GeV}$ on right. On left, dashed and solid lines in blue or brown are for charged and neutral mesons, respectively. Branching ratios are independent of $\xi_1$. }
  \label{BR}
\end{figure}

From eqs.~(\ref{sigco}) and (\ref{gluon}), we consider the relevant interactions of the sigma field to $n_L$ light quarks and gluons in the low energy as
\bea
{\cal L}_{\rm QCD}&=&\frac{1}{2}\frac{ \xi_1}{\sqrt{1+\frac{3}{2}\xi^2_1}}\,\frac\chi{M_P}\,  \left(\sum_{i=1}^{n_L}m_i {\bar q}_i q_i  -\frac{b_{3,L} \alpha_S}{8\pi}G_{\mu\nu}G^{\mu\nu} \right) \nonumber \\
&\equiv & \frac{1}{2}\frac{ \xi_1}{\sqrt{1+\frac{3}{2}\xi^2_1}}\,\frac\chi{M_P}\,  \Theta^\mu_\mu
\eea
with $b_{3,L}=11-\frac{2}{3}n_L$.
Then, using PCAC relation, $\langle\pi^+(p_1)\pi^-(p_2)|\Theta^\mu_\mu|0\rangle=(p_1+ p_2)^2+2m^2_\pi =2p_1\cdot p_2+4m^2_\pi$, we obtain the linear couplings of the sigma field to mesons as
\bea
{\cal L}_{\rm ChPT} 
= \frac{1}{2}\frac{ \xi_1}{\sqrt{1+\frac{3}{2}\xi^2_1}}\,\frac\chi{M_P} \bigg(-(\partial_\mu\pi)^2 +2m^2_{\pi} \pi^2 \bigg), 
\eea
which is nothing but the coupling to the trace of energy-momentum tensor for mesons.
Thus, the Feynman rule for the vertex with one sigma field and two pions with outgoing momenta, $p_1$ and $p_2$, is given by
\bea
V_{\rm \chi\pi\pi}=   \frac{i}{M_P}\frac{ \xi_1}{\sqrt{1+\frac{3}{2}\xi^2_1}} \, \Big(p_1\cdot p_2+2m^2_\pi\Big).
\eea
For instance, for the decay of the sigma field into a pair of pions, we get $p_1\cdot p_2=\frac{1}{2}m^2_\chi-m^2_\pi$.

In Fig.~\ref{BR}, we show the decay branching ratios of the inflaton in the cases of light inflaton below $m_\chi=m_c$ on left and heavy inflaton above $m_\chi=2.5\,{\rm GeV}$ on right. Formulas for inflaton decay rates are collected in the appendix. In the case of light inflaton, the inflaton decays into muons, pions or kaons above the muon threshold while it decays dominantly into an electron pair below the muon threshold but above the electron threshold.
On the other hand, in the case of heavy inflaton, the inflaton decays dominantly into gluons or fermion pairs below the $WW$ threshold, while it decays dominantly into the electroweak sector, $hh, ZZ, WW$, above the $WW$ threshold.

\section{Long-lived inflaton as dark matter}

We consider the sigma field or inflaton as a decaying dark matter and show the parameter space for the correct relic density of the long-lived dark matter, based on Feebly Interacting Massive Particle (FIMP) process after reheating as well as the decays of the inflaton condensate during reheating.

\subsection{Long-lived inflaton}

\begin{figure}
  \begin{center}
    \includegraphics[height=0.42\textwidth]{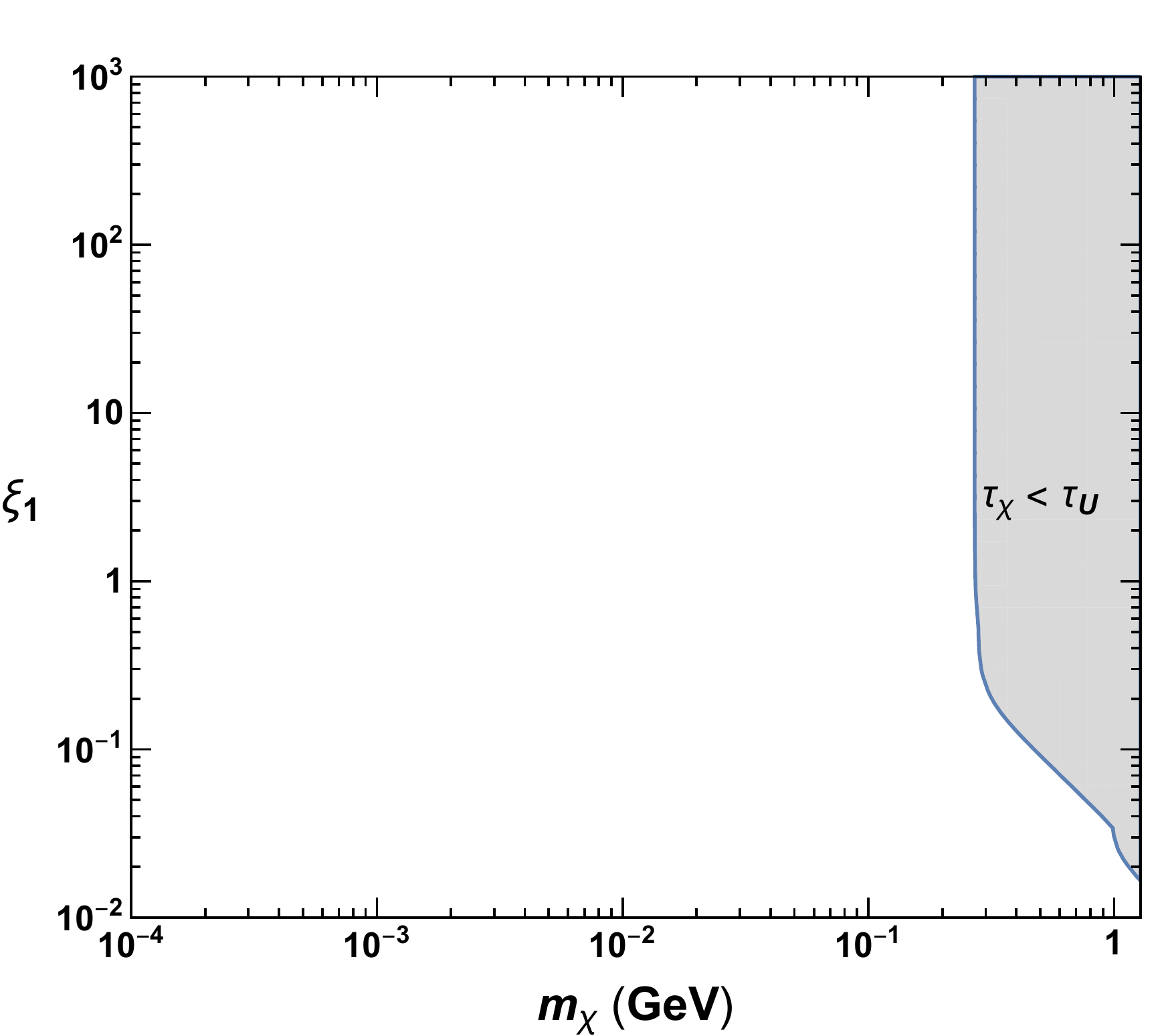}\,\,\,
     \includegraphics[height=0.42\textwidth]{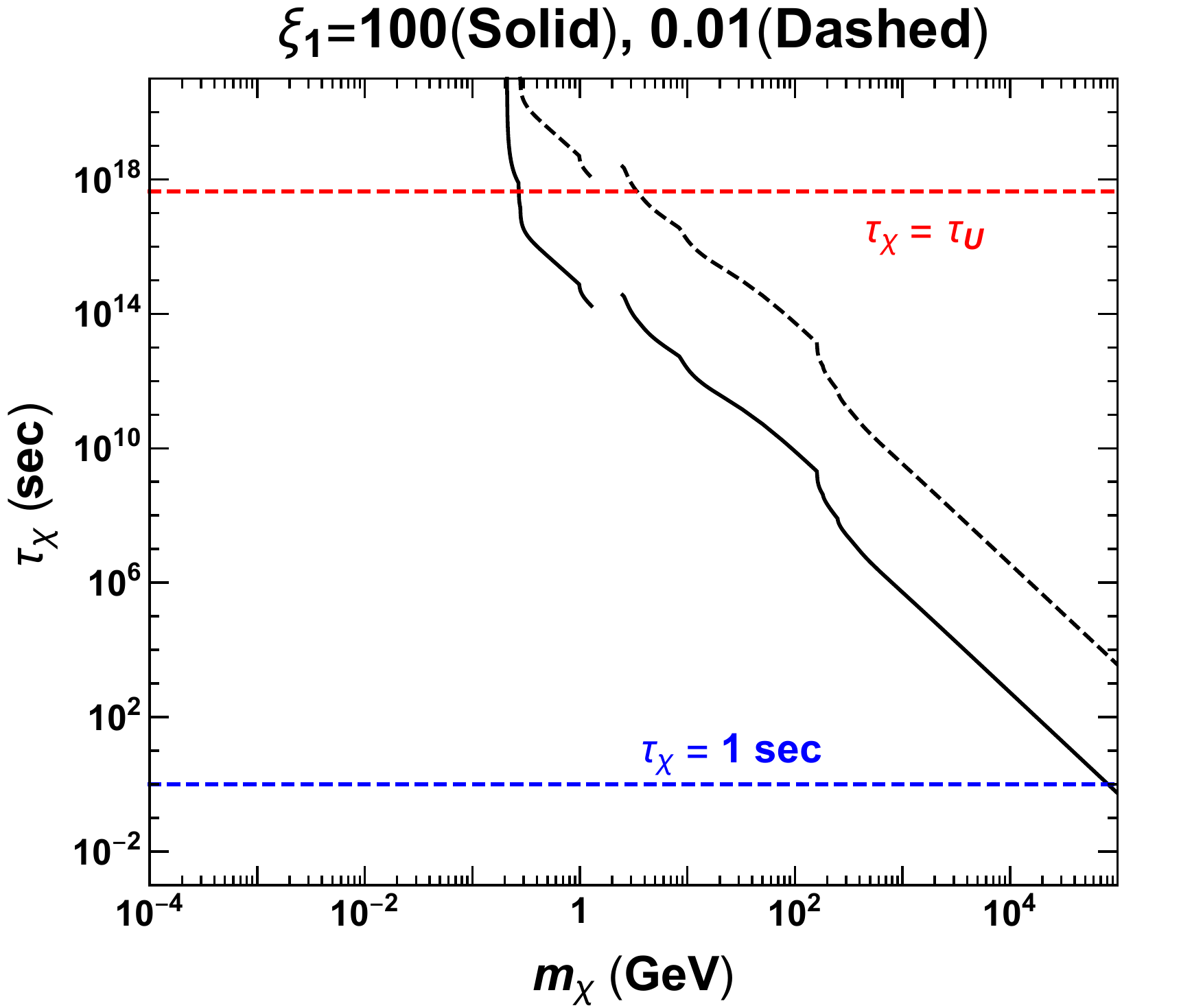}
  \end{center}
  \caption{Left: Parameter space for long-lived inflaton in the plane of $m_\chi$ and $\xi_1$. The lifetime of the sigma field inflaton is shorter than the age of the Universe in gray region. Right: Lifetime of the inflaton as a function of $m_\chi$ for $\xi_1=100, 0.01$ in black solid and dashed lines, respectively. Red and blue dashed lines correspond to $\tau_\chi=\tau_U$ and $1\,{\rm sec}$.}
  \label{lifetime}
\end{figure}

As soon as the decay of the sigma field into a pion pair opens up, the lifetime of the sigma field would be less than the age of the Universe, independent of $\xi_1$ for $\xi_1\gtrsim 1$. Therefore, in most of the parameter space, the sigma field can be a candidate for dark matter only for $m_\chi\lesssim 270\,{\rm MeV}$ \cite{ibarra}. This fact is shown on the left of Fig.~\ref{lifetime}, in the gray region of the parameter space for $m_\chi$ vs $\xi_1$ where the inflaton does not survive until the present Universe.
On the right of Fig.~\ref{lifetime}, we also draw the contours of the inflaton lifetime as a function of $m_\chi$ for $\xi_1=100, 0.01$ in black solid and dashed lines, respectively. We find that the inflaton lifetime ranges between the age of the Universe and $1\,{\rm sec}$ for $m_\chi\approx 270\,{\rm MeV}-10^5\,{\rm GeV}$ with $\xi_1=100$, as shown from the lines with $\tau_\chi=\tau_U$ and $1\,{\rm sec}$.

Dark matter can be in thermal equilibrium, as far as $\lambda_{\chi H}\gtrsim 10^{-7}$ or $\lambda_{\sigma H}\sim \xi^2_1 \lambda_{\chi H}\gtrsim 10^{-3}$ for $\xi_1\sim \sqrt{\xi_2}\sim 10^2$.
But, dark matter can annihilate into a pair of muons or electrons for $m_\chi\lesssim 270\,{\rm MeV}$. For instance, the cross section for the $2\rightarrow 2$ annihilation, $\chi\chi\rightarrow \mu {\bar \mu}$, is suppressed by the SM Higgs mass and the muon Yukawa coupling, as follows,
\bea
\langle\sigma v\rangle_{\chi\chi\rightarrow \mu {\bar \mu}}=  \frac{\lambda^2_{\chi H}m^2_\mu}{4\pi (4m^2_\chi-m^2_h)^2}\, \Big(1-\frac{m^2_\mu}{m^2_\chi} \Big)^{3/2}.
\eea 
On the other hand, the necessary annihilation cross section for thermal freeze-out is $\langle \sigma v\rangle=\frac{\alpha^2_{\rm eff}}{m^2_\chi}$ with the effective DM coupling being given by $\alpha_{\rm eff}\sim 5\times 10^{-6}$ for $m_\chi\sim 100\,{\rm MeV}$.  However, this condition is not satisfied in our model, so we need to rely on non-thermal production mechanisms.

\subsection{Relic density from FIMP inflaton}

For a small mixing quartic coupling between the sigma field and Higgs boson, i.e. $\lambda_{\chi H}\lesssim 10^{-7}$, the sigma field could never be in thermal equilibrium. Thus, the total relic density of inflaton dark matter is determined by two non-thermal mechanisms, as follows,
\bea
\Omega_\chi h^2 =(\Omega_\chi h^2)_{\rm FIMP} +(\Omega_\chi h^2)_{\rm RH}.
\eea
One is the FIMP contribution $(\Omega h^2)_{\rm FIMP} $ \cite{FIMP}, generated by Higgs decay at the temperature $T\gtrsim m_h$. The other is the contribution $(\Omega h^2)_{\rm RH}$ from the decay of the inflaton condensate  during reheating \cite{quarticpot}.

First, in the presence of a nonzero $\lambda_{\chi H}$, the Higgs decay into a pair of sigma fields governs the DM relic density dominantly below the reheating temperature, as follows,
\bea
{\dot n}_\chi+ 3Hn_\chi=2\Big( \Gamma_{h\rightarrow \chi\chi} \, n^{\rm eq}_h-\Gamma_{\chi \chi\rightarrow h} \, n^2_\chi\Big) 
\label{Boltz}
\eea
where the Higgs decay rate is given by
\be
\Gamma_{h\rightarrow \chi\chi} =\frac{\lambda^2_{\chi H}v^2}{32\pi m_h}\sqrt{1-\frac{4m^2_\chi}{m^2_h}},
\ee
the equilibrium number density of Higgs is $n^{\rm eq}_h=\frac{m^3_h}{(2\pi)^{3/2}}\,x^{-3/2}\, e^{-x}$ with $x=m_h/T$ in the non-relativistic limit,
and the second term on right is the inverse decay term, which can be neglected for a small initial abundance of dark matter. 
Then, for $T_{\rm RH}>m_h$, eq.~(\ref{Boltz}) can be solved for $Y_\chi\equiv \frac{n_\chi}{s}$ as
\bea
Y_\chi&=&\frac{2\Gamma_{h\rightarrow\chi\chi}}{H(m_h)s(m_h)} \int^x_{x_{\rm RH}} dx\, x^4 n^{\rm eq}_h  \nonumber \\
&\approx &\frac{0.88\Gamma_{h\rightarrow\chi\chi}M_P }{g^{1/2}_* g_{*s}m^2_h} \int^\infty_0 dx\, x^{5/2} e^{-x}  \nonumber \\
&=&\frac{C\, \Gamma_{h\rightarrow\chi\chi}M_P }{g^{1/2}_* g_{*s}m^2_h}
\eea
with $x_{RH}=m_h/T_{\rm RH}$ and $C=2.9$, which agrees well with the result $C=3.3$ from the exact thermal average \cite{FIMP}. Therefore, the relic density coming from the FIMP process is given by
\bea
(\Omega_\chi h^2)_{\rm FIMP}&=&  2.7\times 10^8 \, Y_\chi\Big(\frac{m_\chi}{1\,{\rm GeV}} \Big) \nonumber \\
&=&0.12 \left( \frac{100}{g_*(m_h)}\right)^{3/2}\left(\frac{\lambda_{\chi H}}{4.4\times 10^{-7}} \right)^2 \left(\frac{m_\chi}{1\,{\rm eV}} \right). \label{FIMP} 
\eea

Next we consider the relic density of inflaton dark matter produced from the  decay of the inflaton condensate during reheating. The energy density of dark matter at the decoupling is given by
\bea
\rho_\chi(a_{\rm dec})= {\rm BR}\cdot \rho_{\chi_c}(a_{\rm dec}) \label{rho1}
\eea
where BR is the branching ratio of the inflaton condensate decaying into a pair of inflatons in eq.~(\ref{BRchi}).
Then, at the decoupling, dark matter has the peak energy at $k=\sqrt{3\lambda_\chi}\chi_0(t_{\rm dec})$ and it becomes non-relativistic when $k(\frac{a_{\rm dec}}{a_{\rm NR}})\sim m_\chi$ at $a=a_{\rm NR}$ due to the redshift of the momentum. Assuming that dark matter becomes non-relativistic before matter-radiation equality for structure formation, the energy density of dark matter at matter-radiation equality is given by
\bea
\rho_\chi(a_{\rm eq})= \rho_\chi(a_{\rm dec}) \Big(\frac{a_{\rm dec}}{a_{\rm NR}} \Big)^4 \Big(\frac{a_{\rm NR}}{a_{\rm eq}} \Big)^3=  \rho_\chi(a_{\rm dec}) \Big(\frac{a_{\rm dec}}{a_{\rm NR}}\Big)\Big(\frac{a_{\rm dec}}{a_{\rm eq}} \Big)^3. \label{rho2}
\eea

First, using eq.~(\ref{decouple}), we obtain the red-shift factor at the time when dark matter becomes non-relativistic as
\bea
\frac{a_{\rm dec}}{a_{\rm NR}}&\sim& \frac{m_\chi}{k}= \frac{m_\chi}{\sqrt{3\lambda_\chi}\chi_0(t_{\rm dec})}=\Big(108\lambda_\chi\Big)^{-1/4} (H_{\rm dec}M_P)^{-1/2} m_\chi. \label{ratio1}
\eea
Then, assuming that there is no entropy change between decoupling and matter-radiation equality, we also get
\bea
\left(\frac{a_{\rm dec}}{a_{\rm eq}} \right)^3= g_{*s}(a_{\rm eq}) (g_*(a_{\rm dec}))^{-1/4} (g_*(a_{\rm eq}))^{-3/4} \left(\frac{H_{\rm eq}}{H_{\rm dec}} \right)^{3/2} \label{ratio2}
\eea
where $g_*(a_{\rm eq})=3.363$, $g_{*s}(a_{\rm eq})=3.909$, $g_*(a_{\rm dec})=106.75$, and $H_{\rm eq}=1.15\times 10^{-37}\,{\rm GeV}$.
Therefore, using the above results and $ \rho_{\chi_c}(a_{\rm dec})=3 H^2_{\rm dec}M^2_P$, we obtain eq.~(\ref{rho2}) with eq.~(\ref{rho1}) explicitly as
\bea
\rho_\chi(a_{\rm eq})=\Big(6.75 \times 10^{-38} \, {\rm GeV}^4\Big) \lambda^{-1/4}_\chi\cdot {\rm BR}\cdot \Big( \frac{m_\chi}{1\,{\rm eV}}\Big). 
\eea
Consequently, we get the general formula for the relic density coming from the reheating process as
\bea
(\Omega_\chi h^2)_{\rm RH}&=&\frac{\rho_\chi(a_{\rm eq})}{\rho_c/h^2}\Big(\frac{a_{\rm eq}}{a_0} \Big)^3 \nonumber \\
&=&0.035\, \lambda^{-1/4}_\chi\cdot{\rm BR}\cdot \Big( \frac{m_\chi}{1\,{\rm eV}}\Big) \nonumber \\
&=&7.3 \, R\, \Big(\frac{r}{0.01} \Big)^{-1/4}\cdot{\rm BR}\cdot \Big( \frac{m_\chi}{1\,{\rm eV}}\Big)
 \label{RHc}
\eea
where the critical density at present is given by $\rho_c=8.05\times 10^{-47}h^2\,{\rm GeV}^4 $, $a_0/a_{\rm eq}=2890$, and in the last line, we used eq.~(\ref{cobe}) and $\lambda_\chi\equiv 4\lambda_\sigma/(9\xi^4_1)=4\lambda_\sigma/(9 \xi^2_2 R^4)$.
In the case that inflation reheats the SM particles dominantly, i.e. ${\rm BR}\approx 11.5 \lambda^2_\chi/\lambda^2_{\chi H}\ll 1$, the above relic density becomes
\bea
(\Omega_\chi h^2)_{\rm RH}&\approx&0.40\,  \lambda^{7/4}_\chi \lambda^{-2}_{\chi H} \left(\frac{m_\chi}{1\,{\rm eV}}\right) \nonumber \\
&=& 0.12 \left(\frac{1.4\times 10^{-8}} {\lambda_{\chi H}}\right)^2R^{-7}\Big(\frac{r}{0.01} \Big)^{7/4}  \left(\frac{m_\chi}{1\,{\rm eV}}\right). \label{RHc-approx}
\eea

Furthermore, from eqs.~(\ref{ratio1}) and (\ref{ratio2}), we obtain the temperature ratios of $T_{\rm eq}$ at matter-radiation equality to $T_{\rm NR}$ at which dark matter becomes non-relativistic, as follows,
\bea
\frac{a_{\rm eq}}{a_{\rm NR}}=\frac{T_{\rm NR}}{T_{\rm eq}}&=& 0.77 \lambda^{-1/4}_\chi \Big(\frac{m_\chi}{1\,{\rm eV}} \Big) \nonumber \\
 &=&160\, R\, \Big(\frac{r}{0.01} \Big)^{-1/4} \Big(\frac{m_\chi}{1\,{\rm eV}} \Big). \label{TNR}
\eea
Here, for $R={\cal O}(1)$ and $r=0.01$, we find that $T_{\rm NR}$ is greater than $T_{\rm BBN}$ for $m_\chi>7.8\,{\rm keV}$, which is not favored by the correct relic density, as will be discussed shortly. 

In the case with $T_{\rm NR}<T_{\rm BBN}$, dark matter is still relativistic during BBN, so we need to check the contribution of dark matter to the number of relativistic species, $\Delta N_{\rm eff}$. 
Assuming that dark matter is still relativistic during BBN and using eq.~(\ref{rho2}), we get the DM relic density for $a>a_{\rm BBN}$ as
\bea
\rho_\chi(a)&=&\rho_\chi(a_{\rm dec}) \left(\frac{a_{\rm dec}}{a} \right)^4 \nonumber \\
&=& \rho_\chi(a_{\rm eq}) \left(\frac{a_{\rm NR}}{a_{\rm eq}} \right) \left(\frac{a_{\rm eq}}{a} \right)^4  \nonumber \\
&=& \frac{ \rho_\chi(a_{\rm eq}) }{\rho_R(a_{\rm eq})}\,\left(\frac{a_{\rm NR}}{a_{\rm eq}} \right)  \rho_R(a).
\eea
Then, from $\rho_R(a)=\frac{\pi^2}{30}\, g_* T^4$ and $\Delta\rho=\frac{\pi^2}{30}\cdot \frac{7}{4}\Big(\frac{4}{11}\Big)^{4/3}(\Delta N_{\rm eff})\, T^4$, we obtain $\Delta N_{\rm eff}$ from dark matter during BBN as follows,
\bea
\Delta N_{\rm eff} &=& \frac{4}{7}\Big(\frac{11}{4}\Big)^{4/3} \, g_*\,\cdot   \frac{ \rho_\chi(a_{\rm eq}) }{\rho_R(a_{\rm eq})}\,\cdot\left(\frac{a_{\rm NR}}{a_{\rm eq}} \right)  \nonumber \\
&\leq& 0.0944 \, R^{-1}  \Big(\frac{r}{0.01} \Big)^{1/4} \left(\frac{1\,{\rm eV}}{m_\chi}\right)
\eea
where the inequality comes from $ \rho_\chi(a_{\rm eq})\leq\rho_{\rm DM}(a_{\rm eq})=\rho_R(a_{\rm eq})$, and we took $g_*=6.863$ for $0.5\,{\rm MeV}<T<1\,{\rm MeV}$. 
The combined results of  primordial abundance measurements of helium and deuterium and the CMB measurement by Planck constrain $\Delta N_{\rm eff}$ to be $-0.116\pm 0.23$ in case a),  $-0.006\pm 0.22$ in case b), or $0.014\pm 0.22$ in case c), depending on the computed deuterium fraction \cite{planck2018}.
Therefore, our inflaton dark matter is consistent with such BBN constraints, as far as $m_\chi\gtrsim 0.139(0.104)\,{\rm eV}$ within $2\sigma$ in case c), for $ \rho_\chi(a_{\rm eq})=\rho_{\rm DM}(a_{\rm eq})$, $R=1.5(2)$ and $r=0.01$.

 \begin{figure}
  \begin{center}
    \includegraphics[height=0.42\textwidth]{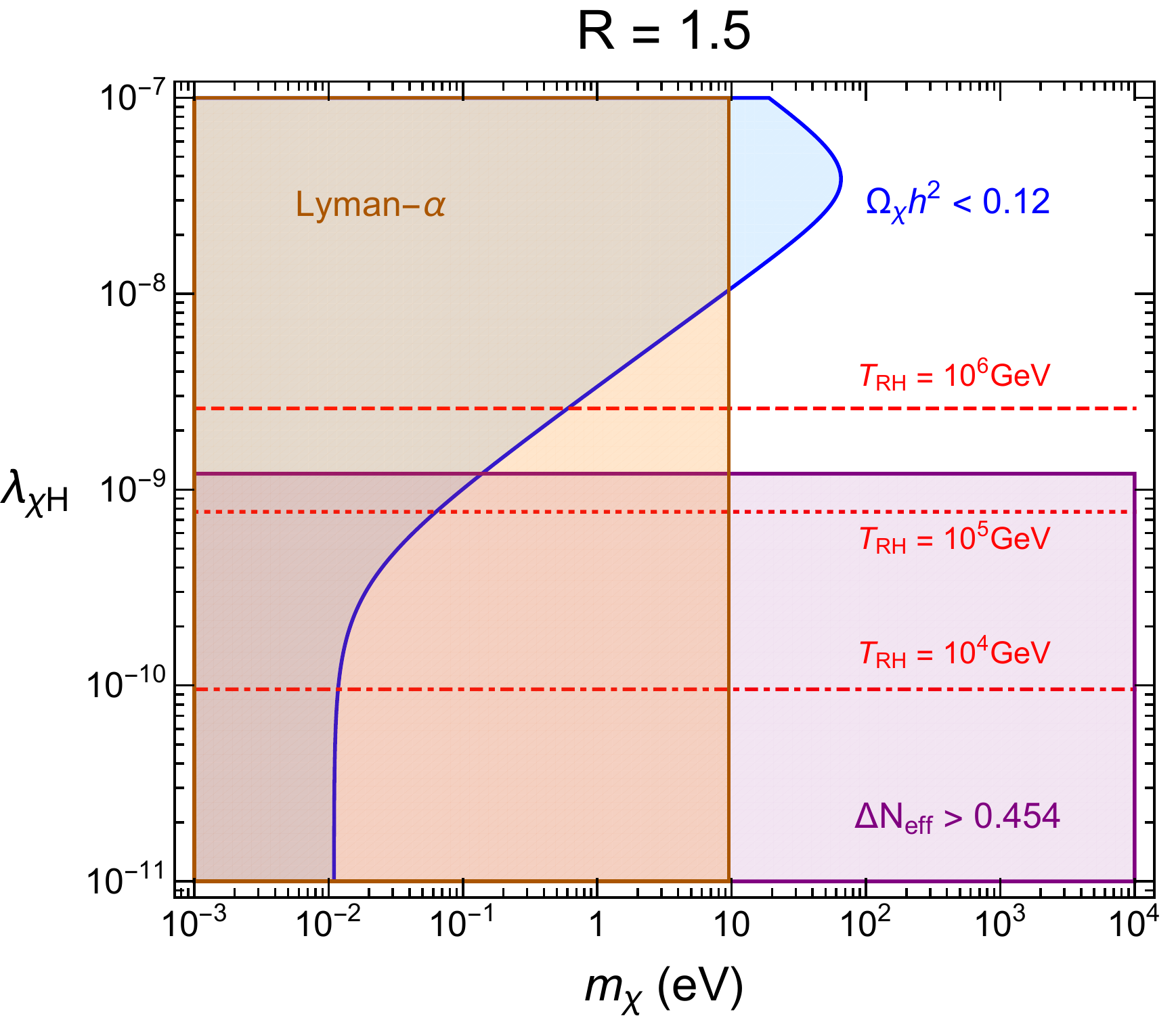}\,\,\,\,\,
      \includegraphics[height=0.42\textwidth]{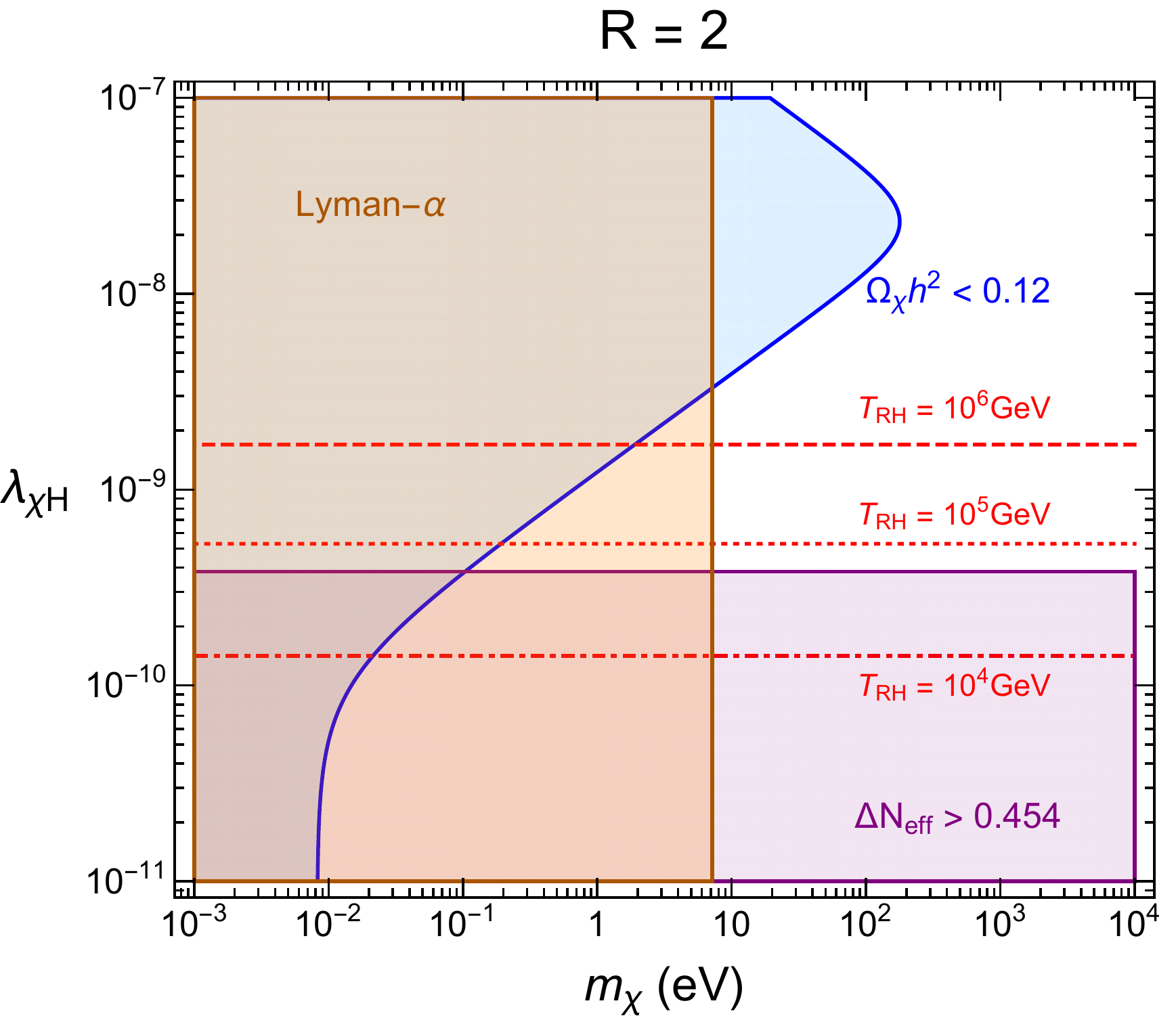}
  \end{center}
  \caption{Parameter space for the DM relic density produced by non-thermal mechanisms in the plane of $m_\chi$ and $\lambda_{\chi H}$.  The relic density satisfies $\Omega_\chi h^2<0.12$ in light blue region while the light orange region is ruled out by the Lyman-$\alpha$ forest and phase space density bounds.
In purple region, dark matter contributes to $\Delta N_{\rm eff}>0.454$ during BBN. Contours with reheating temperature, $T_{\rm RH}=10^6, 10^5, 10^4\,{\rm GeV}$, are shown on left(right) in red dashed, dotted, dot-dashed, and solid lines, respectively. $R=\xi_1/\sqrt{\xi_2}=1.5, 2$ is chosen on left and right plots, respectively, and $r=0.01$ is taken for both plots. }
  \label{relic}
\end{figure}

In Fig.~\ref{relic}, we show the parameter space for the DM relic density due to non-thermal production mechanisms in light blue region for $m_\chi$ and $\lambda_{\chi H}$, for $R=\xi_1/\sqrt{\xi_2}=1.5, 2$, on left and right plots, respectively. Using the reheating temperature obtained in eq.~(\ref{TRH}), we also indicate the contours with reheating temperature, $T_{\rm RH}=10^6, 10^5, 10^4\,{\rm GeV}$, on left(right), in red dashed, dotted, dot-dashed (and solid) lines, respectively.
Moreover, the light orange region is excluded by the bounds on the free-streaming length up to $\lambda_{\rm fs}<0.6\,{\rm Mpc}$, from the Lyman-$\alpha$ forest data \cite{lyman} and the phase space densities derived from the dwarf galaxies of the Milky way \cite{dwarf}.
Finally, purple region is with $\Delta N_{\rm eff}>0.454$, which is beyond the $2\sigma$ limit of the BBN constraint in case c). 

As a result,  we find that light dark matter with $7\,{\rm eV}\lesssim m_\chi\lesssim 200\,{\rm eV}$ is favored for the correct relic density, being compatible with BBN and CMB constraints.  As the decay branching fraction ${\rm BR}$ of the inflaton condensate into an inflaton pair gets larger, the relic density becomes almost independent of the inflaton-Higgs quartic coupling, $\lambda_{\chi H}$, and the reheating temperature gets smaller. But, the region with a large ${\rm BR}$ is disfavored by BBN constraints.
On the other hand, for $\lambda_{\chi H}\gtrsim 10^{-8} (3\times 10^{-9})$ for $R=1.5(2)$, the inflaton condensate decays dominantly into a Higgs pair, so the relic density is saturated along the line with constant $m_\chi/\lambda^2_{\chi H}$, as expected from the approximate formula in eq.~(\ref{RHc-approx}).

We remark briefly on other potential constraints on the inflaton dark matter.
We note that there is no mixing between Higgs and sigma fields in our model so there is no direct constraint on the mixing quartic coupling, $\lambda_{\chi H}$, in the relevant parameter space for the correct relic density, and indirect constraint from Higgs invisible decays into a pair of sigma fields is not sensitive enough to bound such a tiny coupling. 
Furthermore, there are couplings of the sigma field to photons through the trace of the energy-momentum tensor in eq.~(\ref{photon}) but such couplings are suppressed by the Planck scale, so there is no constraint from SN1987A or horizontal branch cooling \cite{starcool} or fifth-force experiments \cite{fifth}. On the other hand, the bounds from isotropic diffuse gamma-ray spectrum and CMB measurements \cite{ibarra} can constrain the parameter space for a decaying dark matter heavier than $m_\chi\sim 2\,{\rm MeV}$, but there is no constraint in the parameter space for FIMP dark matter in our model.
There are also interesting constraints by electron absorption from XENON10 \cite{xenon10} on the detection of a light dark matter below $10\,{\rm eV}$ or proposed experiments with superconductors or semi-conductors \cite{detect10eV}, but the sensitivity has not reached yet to probe our inflaton dark matter.

\section{Conclusions}

We have studied the dynamics of inflation models of a singlet scalar field with both quadratic and linear non-minimal couplings. Although the quadratic non-minimal coupling determines the flat direction for inflation, the linear non-minimal coupling starts to dominate already during reheating and rescales the effective quartic couplings and mass of the inflaton to small values. We identified the reheating temperature in this model and obtained the correct abundance of the inflaton dark matter by non-thermal production mechanisms with the decay of the inflaton condensate during reheating and the decay of Higgs after reheating. 

It is intriguing that the inflaton couples to the trace of the energy-momentum tensor so does it to the full Jordan frame potential. As a result, there is no mixing between the inflaton and the SM Higgs boson in the vacuum, allowing for a definite prediction for the inflaton decay rates in terms of the linear non-minimal coupling and the inflaton mass. 
We showed that the effective quartic coupling of the inflaton is fixed by the CMB normalization while a tiny mixing quartic coupling between the inflaton and the SM Higgs boson can be varied to saturate the relic density for DM masses between $m_\chi\sim 7\,{\rm eV}$ and $200\,{\rm eV}$, being compatible with BBN, CMB and large-scale structure constraints.

\section*{Acknowledgments}

We thank Fedor Bezrukov, Cristiano Germani, Pak Hang Chris Lau, Wan-Il Park and Chang Sub Shin for their valuable comments and discussions. 
HML appreciates fruitful discussions with participants during the CERN-CKC Theory Workshop on Scale Invariance in Particle Physics and Cosmology.
The work of SMC, HML and YJK is supported in part by Basic Science Research Program through the National Research Foundation of Korea (NRF) funded by the Ministry of Education, Science and Technology (NRF-2016R1A2B4008759 and NRF-2018R1A4A1025334). 
The work of KY is supported in part by National Center for Theoretical Sciences.
The work of SMC is supported in part by TJ Park Science Fellowship of POSCO TJ Park Foundation.
The work of YJK is supported in part by the Chung-Ang University Graduate Research Scholarship in 2018.

\def\theequation{A.\arabic{equation}}

\setcounter{equation}{0}

\vskip0.8cm
\noindent
{\Large \bf Appendix A: Inflaton decay rates} 
\vskip0.4cm
\noindent

The sigma inflaton field has couplings to the SM particles through the trace of the energy-momentum tensor. 
Here, we list formulas for the most relevant two-body decay rates of the inflaton, as follows \cite{diphoton,radion,radion1},
\bea
\Gamma(hh)&=&\frac{|V_{ \chi hh}|^2}{32\pi m_\chi } 
\sqrt{1-\frac{4m^2_h}{m^2_\chi } } \nonumber \\ 
&=& \frac{(2m^2_h+m^2_\chi)^2}{128\pi m_\chi M^2_P}\left(\frac{ \xi^2_1}{1+\frac{3}{2}\xi^2_1} \right) \sqrt{1-\frac{4m^2_h}{m^2_{\chi } } }, \\
\Gamma({\bar f}f) &=&\frac{g^2_{\chi ff} m_{\chi }}{8\pi}\, \Big( 1-\frac{4m^2_f}{m^2_{\chi } }\Big)^{3/2} \nonumber \\
&=& \frac{m^2_f m_{\chi }}{32\pi M^2_P} \left(\frac{ \xi^2_1}{1+\frac{3}{2}\xi^2_1} \right) \Big( 1-\frac{4m^2_f}{m^2_{\chi } }\Big)^{3/2}, \\
\Gamma(VV)&=& \frac{\delta_V g_{\chi VV}^2 m^3_\chi}{32\pi m^4_V}\Big( 1-4\frac{m^2_V}{ m^2_{\chi } }+12\frac{m^4_V}{ m^4_{\chi } } \Big)\sqrt{1-\frac{4m^2_V}{m^2_{\chi }  }} \nonumber  \\
&=& \frac{\delta_V m^3_\chi }{128\pi M^2_P} \left(\frac{ \xi^2_1}{1+\frac{3}{2}\xi^2_1} \right)\Big( 1-4\frac{m^2_V}{ m^2_{\chi } }+12\frac{m^4_V}{ m^4_{\chi } } \Big)\sqrt{1-\frac{4m^2_V}{m^2_{\chi }  }}, \\ 
\Gamma(\gamma\gamma) &=& \frac{g^2_{\chi\gamma\gamma} }{4\pi}\, m^3_\chi \nonumber  \\
&=& \frac{ \alpha^2  }{1024\pi^2}  \frac{ m^3_\chi }{M^2_P}  \left(\frac{ \xi^2_1}{1+\frac{3}{2}\xi^2_1} \right)\left|b_2+b_Y +A_W(x_W)+2\sum_{f=q,l} N_c Q^2_f A_F(x_f) \right|^2,  \label{decaygaga} \\
\Gamma(gg)&=& \frac{2g^2_{\chi gg} }{\pi}\, m^3_\chi \nonumber  \\
&=& \frac{ \alpha^2_S  }{128\pi^2} \frac{ m^3_\chi} {M^2_P} \left(\frac{ \xi^2_1}{1+\frac{3}{2}\xi^2_1} \right)\left|b_3+ \sum_{f=q} A_F(x_f)\right|^2. \label{decaygg}
\eea
Here,  $\delta_V=1(2)$ for $V=Z(W)$ bosons, $b_Y, b_2,  b_3$ are the beta function coefficients of  $U(1)_Y$, $SU(2)_L$ and $SU(3)_C$ gauge couplings, given by $(b_Y, b_2,  b_3)=(-\frac{41}{6},\frac{19}{6},7)$ in the SM, leading to $b_\gamma=b_2+b_Y=-\frac{11}{3}$ for the beta function of EM gauge coupling, $x_f=4m^2_f/m^2_\chi$, $x_W=4m^2_W/m^2_\chi$, and  the loop functions are given by
\bea
A_F(x)&=&x\Big(1+ (1-x)f(x) \Big) \\
A_W(x)&=&-\Big(2+3x+3x(2-x)f(x)\Big)
\eea
where
\bea
f(x)=\left\{ \begin{array}{cc} {\rm arcsin}^2\Big(\frac{1}{\sqrt{x}}\Big), \quad \quad x\geq1 \\
-\frac{1}{4} \Big[{\rm log}\frac{1+\sqrt{1-x}}{1-\sqrt{1-x}}-i\pi\Big]^2, \quad x<1. \end{array} \right.
\eea
In the limit of decoupled particles, the loop functions are approximated to $A_F(x)\rightarrow \frac{2}{3}$ and $A_W(x)\rightarrow -7$ for $x\gg 1$, thus recovering the low energy couplings coming from trace anomalies due to light particles only: $b_2+b_Y+A_W(x_W)+2\sum_{f=q,l} N_c Q^2_fA_F(x_f) \rightarrow b_{\gamma,L}$ in eq.~(\ref{photon}) and $b_3+ \sum_{f=q} A_F(x_f)\rightarrow b_{3,L}$ in eq.~(\ref{gluon}). 

For simplicity, we took the notations, $m_f, m_V, m_h$, for the SM particle masses that are independent of the inflaton field value.  

For $m_\chi<1.5\,{\rm GeV}$, we need to rely on chiral perturbation theory to obtain the  decay rates of the inflaton into a meson pair, as follows \cite{radion1},
\bea
\Gamma(\pi^a\pi^a) &=&\frac{|V_{\rm \chi\pi\pi}|^2}{32\pi m_\chi } 
\sqrt{1-\frac{4m^2_\pi}{m^2_\chi } } \nonumber \\
&=& \frac{(2 m^2_\pi+ m_\chi^2 )^2}{128\pi m_\chi M^2_P}\left(\frac{ \xi^2_1}{1+\frac{3}{2}\xi^2_1}\right) 
\sqrt{1-\frac{4m^2_\pi}{m^2_{\chi } } }. 
\eea

\end{document}